\documentclass[fleqn,usenatbib]{mnras}

\usepackage{newtxtext,newtxmath}
\usepackage[T1]{fontenc}
\usepackage{comment}

\DeclareRobustCommand{\VAN}[3]{#2}
\let\VANthebibliography\thebibliography
\def\thebibliography{\DeclareRobustCommand{\VAN}[3]{##3}\VANthebibliography}

\usepackage{graphicx}	% Including figure files
\usepackage{amsmath}	% Advanced maths commands

\title[A pebble CPD around a low-mass planet]{A Circumplanetary Pebble Disk around a Three-Earth-Mass Protoplanet in a Wind-Launching Non-ideal MHD Disk}

\author[C.-Y. Hsu et al.]{
Chun-Yen Hsu,$^{1,2}$\thanks{E-mail: kdj8qp@virginia.edu}
Zhi-Yun Li,$^{1,2}$
Xiao Hu,$^{3}$
Yisheng Tu,$^{4}$
Min-Kai Lin,$^{5,6}$
\\
$^{1}$Department of Astronomy, University of Virginia, Charlottesville, VA 22904, USA\\
$^{2}$ Virginia Institute of Theoretical Astronomy, University of Virginia, Charlottesville, VA 22904, USA\\
$^{3}$Department of Astronomy, University of Florida, Gainesville, FL 32608, USA\\
$^{4}$ Department of Astronomy, University of Michigan, Ann Arbor, MI, 48109, USA\\
$^{5}$Institute of Astronomy and Astrophysics, Academia Sinica, Taipei 10617, Taiwan\\
$^{6}$Physics Division, National Center for Theoretical Sciences, Taipei 10617, Taiwan \\
}

\date{Accepted XXX. Received YYY; in original form ZZZ}

\pubyear{\the\year{}}

\begin{document}
\label{firstpage}
\pagerange{\pageref{firstpage}--\pageref{lastpage}}
\maketitle


\begin{abstract}
Pebble accretion can rapidly build planetary cores, but its rate depends on how much incoming solid material reaches the protoplanet. It is unclear whether pebbles entering the Hill sphere of a subthermal protoplanet promptly reach its surface or accumulate in their own disk within the pressure-supported gas envelope. We investigate this with a global three-dimensional non-ideal magnetohydrodynamic simulation of a wind-launching disk hosting a three-Earth-mass (subthermal) protoplanet at 10~au, following gas and dust from the circumstellar disk into the Hill sphere, including dust feedback. We find that centimeter-sized pebbles form a thin, prograde circumplanetary disk (pebble CPD), even though the planet opens no deep gap and its gas envelope remains approximately spherical. The local pebble-to-gas density ratio rises up to $10^4$ times its initial value, making the CPD dust-dominated, with an average pebble-to-gas mass ratio of order 10 or more. Radial force profiles and small mean radial velocities show that the CPD is primarily rotationally supported in the inner disk, with gas drag contributing farther out. Horizontal transport and vertical settling first build the reservoir. After slower accumulation and intermittent losses, growth resumes with enhanced vertical feeding, consistent with gas-advection-assisted delivery. The pebble mass reaches approximately $4\times10^{-2}\,M_\oplus$ in approximately $10^3$~yr. Dust feedback may sustain the disk by bringing the gas into near-corotation with the pebbles, slowing their angular-momentum loss through aerodynamic drag. This reservoir adds an intermediate stage between pebble capture and delivery to the protoplanet and could supply material for early satellite formation.

\end{abstract}


%%%%%%%%%%%%%%%%%%%%%%%%%%%%%%%%%%%%%%%%%%%%%%%%%%

%%%%%%%%%%%%%%%%% BODY OF PAPER %%%%%%%%%%%%%%%%%%

\section{Introduction}
\label{sec: Intro}

Pebbles approaching a protoplanet embedded in a protoplanetary disk can pass through its
gravitational sphere of influence, accrete onto the planet, or
remain temporarily in a circumplanetary reservoir. Distinguishing
these outcomes is essential for connecting the radial transport
of solids through protoplanetary disks to planetary growth.
Gas drag can facilitate pebble capture by dissipating encounter
energy and enlarging the effective accretion cross-section
\citep{Ormel10,Lambrechts12}, making pebble accretion a promising
route to the assembly of planetary cores. However, entry into
the Hill sphere, retention in a circumplanetary structure, and
accretion onto a planetary surface describe different stages
of this process. A particularly interesting possibility is that
pebbles form a rotationally supported disk around a low-mass planet even when the surrounding gas forms an approximately spherical envelope. 

This contrast arises because gas and solids respond differently
to planetary gravity. Gas pressure can support an extended
envelope around an embedded low-mass planet, while pebbles have
no comparable thermal pressure support. Their spatial distribution can differ substantially from that of the gas before entering the Hill sphere (due, e.g., to vertical dust settling), and their motion depends instead on gravity, angular momentum, and aerodynamic coupling to the gas. Because of these differences, it is not obvious whether the captured pebbles promptly reach the protoplanet or accumulate in a circumplanetary pebble disk (hereafter "pebble CPD"). The latter raises two related questions: how its constituent solids acquire and retain orbital angular momentum and what dynamical balance allows the disk to persist. We use this term for a flattened concentration of pebbles rotating around the planet, with rotation dominating infall; it does not require a rotationally supported circumplanetary gas disk\footnote{
 Simulations show that the rotational support of planetary gas envelopes depends on planet mass relative to thermal mass, thermodynamics, and the treatment of the inner planetary region \citep{Bethune2019,Fung2019}. Although increasing planet mass generally favors stronger rotation, isothermal calculations can produce compact rotationally supported gas disks even around subthermal planets.}.

Earlier calculations provide important precedents.
\citet{Johansen2010} found that pebble streams
interacting with gas within a protoplanet's Hill sphere can
spiral into a prograde circumplanetary disk and deliver prograde
angular momentum to the accreting body. Their calculations also demonstrated the importance of dust loading and aerodynamic feedback in maintaining prograde particle orbits around relatively small bodies of hundreds of kilometers.
\citet{Ormel2013} showed that a low-mass planet modifies the
surrounding gas flow and that the resulting envelope structure
and encounter geometry affect solid-particle trajectories:
tightly coupled particles can follow gas streamlines away from
the planet, whereas pebbles can spiral in toward it. More recently, \citet{Kuwahara2026} found that
larger particles settle into flattened distributions within
convectively stable envelopes of subthermal planets embedded in an inviscid circumstellar disk, although
radial infall toward the planet dominates their motion over the region resolvable by their calculations. Global multifluid calculations by \citet{Krapp2022}
also reveal vertically stratified dust distributions within
planetary envelopes around protoplanets embedded in a standard $\alpha$-disk, focusing on their implications for opacity. 
These studies motivate us to examine pebble capture from a more actively accreting, wind-launching, non-ideal MHD circumstellar disk and the storage of captured pebbles in circumplanetary space. Specifically, we ask whether a dust-dominated pebble CPD develops and, if so, how it is supported and how it exchanges solids with its evolving surroundings. 

The global circumstellar disk environment matters because it sets the spatial
distribution and velocities of the pebbles reaching the Hill sphere of the protoplanet. 
In nonideal magnetohydrodynamic (MHD) disks, magnetic stresses
and disk winds can transport angular momentum and produce
accretion flows that vary with height and radius
\citep[e.g.,][]{Bai2013a,Suriano17,Bethune17, Riols19,  Cui21,  Lesur2021,  Tabone2022}. Global calculations also
show the development of rings, gaps, and vertically structured
flows associated with magnetic flux redistribution and
wind-driven accretion \citep[e.g.,][]{Hu22,Hsu24}. Dust responds to these
structures through grain-size-dependent settling and radial
drift. In the planet-free calculations of \citet{Hsu25}, small
grains remain closely coupled to the gas, while larger grains
settle more strongly and migrate through the evolving ring-and-gap
structure. Planet--disk--wind interactions introduce additional
spatial structure in the delivery of material to circumplanetary
regions \citep{Aoyama23,WafflardFernandez2023,Hsu26}.
These results motivate following the coupled gas and dust
dynamics from the magnetized circumstellar disk down to the protoplanet's neighborhood rather than prescribing a steady local pebble supply.

Our investigation builds on the global three-dimensional non-ideal MHD
simulations of dust and gas in wind-launching disks with embedded planets
by \citet{Hsu26}. We focus on the case of a subthermal planet with mass
$M_{\rm p}=0.01\,M_{\rm J}$ (or approximately $3\,M_\oplus$) orbiting at 10~au,
corresponding to a dimensionless mass of
$m\equiv M_{\rm p}/M_{\rm th}\simeq0.076$, where $M_{\rm th}$ is the
thermal mass. We improve the resolution in the Hill sphere using
static mesh refinement, so that the dust and gas structures are better
resolved near the planet. We find that centimeter-sized pebbles assemble
into a thin, dust-dominated, prograde circumplanetary disk
(``pebble CPD'' hereafter), while the gas develops an approximately spherical envelope. After describing the numerical model in Section~\ref{sec:simulation}, we trace the gas and dust distributions
from the global disk to the Hill sphere in Section~\ref{sec:results}.
Section~\ref{sec:dust in the Hill sphere} examines what supports the
pebble CPD, how horizontal transport and vertical settling feed it,
and how the balance of supply and loss shapes its mass evolution.
Section~\ref{sec:discussion} places these results in a broader context and discusses their implications for pebble delivery
and satellite formation. Section~\ref{sec:conclusion} summarizes
our findings.

\section{SIMULATION SETUP} 
\label{sec:simulation}

We use Athena++ \citep{stone20} with a pressureless dust fluid module  \citep{Huang22,Huang25} to solve the non-ideal MHD gas and dust fluid equations:
\begin{equation}
  \frac{\partial \rho_{\rm g}}{\partial t} + \nabla \cdot (\rho_{\rm g} {\rm \bf V}_{\rm g}) = 0, \label{eq:gas_continuity}
\end{equation}
\begin{equation}
  \frac{\partial (\rho_{\rm g} {\rm \bf V}_{\rm g})}{\partial t} + \nabla \cdot (\rho_{\rm g} {\rm \bf V}_{\rm g} {\rm \bf V}_{\rm g} + {\rm P^*} {\rm {\bf I}}  - \frac{{\rm \bf B} {\rm \bf B} }{4 \pi}) = - \rho_{\rm g} \nabla \Phi + \rho_{\rm d} \frac{{\rm \bf V}_{\rm d} - {\rm \bf V}_{\rm g}}{\tau_{\rm s}}, \label{eq:gas_momentum}
\end{equation}
\begin{align}
  \frac{\partial E_{\rm g}}{\partial t} + \nabla \cdot  \left[(E_{\rm g} + {\rm P^*}){\rm \bf V}_{\rm g} - \frac{{\rm \bf B} ({\rm \bf B} \cdot {\rm \bf V}_{\rm g})}{4 \pi} + \frac{1}{c} \left( \eta_O {\rm \bf J} + \eta_{\rm AD} {\rm \bf J}_{\perp} \right) \times {\rm \bf B}  \right] \notag \\
  = - \rho_{\rm g} ({\rm \bf V}_{\rm g} \cdot \nabla \Phi) - \Lambda_{\rm c} + \rho_{\rm d} \frac{{\rm \bf V}_{\rm d} - {\rm \bf V}_{\rm g}}{\tau_{\rm s}} \cdot {\rm \bf V}_{\rm g} + \omega_{\rm d} \rho_{\rm d} \frac{|{\rm \bf V}_{\rm d} - {\rm \bf V}_{\rm g}|^2}{\tau_{\rm s}}, \label{eq:gas_energy}
\end{align}
\begin{equation}
\frac{\partial \rho_{\rm d}}{\partial t} + \nabla \cdot (\rho_{\rm d} {\rm \bf V}_{\rm d} ) = 0, \label{eq:dust_continuity}
  %\frac{\partial \rho_{\rm d}}{\partial t} + \nabla \cdot (\rho_{\rm d} {\rm \bf V}_{\rm d} + {\bf \mathcal{F}}_{\rm dif}) = 0, \label{eq:dust_continuity}
\end{equation}
\begin{equation}
\frac{\partial \rho_{\rm d} ({\rm \bf V}_{\rm d})}{\partial t} + \nabla \cdot (\rho_{\rm d} {\rm \bf V}_{\rm d} {\rm \bf V}_{\rm d}) = - \rho_{\rm d} \nabla \Phi + \rho_{\rm d} \frac{{\rm \bf V}_{\rm g} - {\rm \bf V}_{\rm d}}{\tau_{\rm s}}, \label{eq:dust_momentum}
  %\frac{\partial \rho_{\rm d} ({\rm \bf V}_{\rm d} + {\rm \bf V}_{\rm d, dif})}{\partial t} + \nabla \cdot (\rho_{\rm d} {\rm \bf V}_{\rm d} {\rm \bf V}_{\rm d} + {\bf \Pi}_{\rm dif}) = - \rho_{\rm d} \nabla \Phi + \rho_{\rm d} \frac{{\rm \bf V}_{\rm g} - {\rm \bf V}_{\rm d}}{T_{\rm s}}, \label{eq:dust_momentum}
\end{equation}
and the induction equation
\begin{equation}
  \frac{\partial {\rm \bf B}}{\partial t} = \nabla \times ({\rm \bf V}_{\rm g} \times {\rm \bf B}) - \frac{4 \pi}{c} \nabla \times \left[ \eta_O {\rm \bf J} + \eta_{\rm AD} {\rm \bf J}_{\perp} \right] , \label{eq:induction}
\end{equation}
where $\rho_{\rm g}$ and ${\rm \bf V}_{\rm g}$ are gas mass density and velocity, $\rho_{\rm d}$ and ${\rm \bf V}_{\rm d}$ are dust mass density and velocity, ${\rm \bf B}$ is the magnetic field, ${\rm P^*} = P_{\rm g} + B^2/(8\pi)$ is the total (thermal [$P_{\rm g}$] and magnetic)  pressure, ${\rm \bf I}$ is the identity tensor, $\Phi = -GM/r$ is the gravitational potential of the central star, $E_{\rm g} = \rho_{\rm g} V_{\rm g}^2/2 + {P_{\rm g}}/(\gamma - 1) + B^2/(8 \pi)$ is the energy density, $\gamma$ is the adiabatic index, ${\rm \bf J}$ is the current density, ${\rm \bf J}_{\perp} = {\rm \bf B} \times ({\rm \bf J} \times {\rm \bf B}) / (B^2)$ is the component of ${\rm \bf J}$ perpendicular to the magnetic field, $\eta_O$ and $\eta_{\rm AD}$ are the Ohmic and ambipolar diffusivities, and $\Lambda_{\rm c}$ is the cooling term.
The aerodynamic drag between gas and dust is included in the last terms of Eq. \ref{eq:gas_momentum} and \ref{eq:dust_momentum}. The dust fluid module in Athena++ assumes linear drag law, so the stopping time $\tau_{\rm s}$ is independent of velocity.
The last two source terms on the right hand side of the gas energy equation (Eq.~\ref{eq:gas_energy}) are drag and frictional heating terms, respectively. The parameter $\omega_{\rm d}$ is used to control the level of frictional heating, and we set $\omega_{\rm d} = 1$ to assume all the dissipation is deposited in the gas. Although we solve the full energy equation (Eq. 3), we employ a rapid cooling scheme that keeps the disk close to locally isothermal during the evolution. Our simulations are conducted in a star-centered spherical-polar coordinate system rotating at the protoplanet’s orbital angular velocity, so that the protoplanet remains fixed on the grid. We follow the implementation described in Appendix A of \citealt{Aoyama23}. Unless otherwise stated, velocities are reported in this corotating frame.

\subsection{Simulation domain} \label{subsection:simulation_domain}

We perform fully 3D simulations using spherical polar coordinates ($r, \theta, \phi$). The simulation domain spans 1 to 316~au radially, 0.05 to $\pi$-0.05 in the polar direction, and 0 to $2\pi$ azimuthally. The radial grid follows logarithmic spacing with 80 base cells and an adjacent-cell size ratio of 1.0746. The polar grid is uniform with 96 base cells, and the azimuthal grid is uniform with 48 base cells. 
Eight static mesh-refinement levels increase
the resolution by a factor of two each. The two finest levels focus on the
planet at $\phi=\pi$, while coarser refinement covers the circumstellar
disk. Table \ref{tab:refinement} summarizes the spatial extents of refinement.

\begin{table}
\centering
\caption{Spatial extents of refinement. The angular widths are measured from the midplane and the planet's azimuth for $\theta$ and $\phi$, respectively.}
\label{tab:refinement}
\begin{tabular}{cccc}
\hline
Level & $r$ (au) & $|\theta-\pi/2|$ & $|\phi-\pi|$\\
\hline
1 & 3.14--316 & $\leq1.0$ & all\\
2 & 5.57--176 & $\leq0.5$ & all\\
3 & 2.37--100 & $\leq0.25$ & all\\
4 & 7.5--13.33 & $\leq0.126$ & $\leq0.26$\\
5 & 8.66--11.54 & $\leq0.095$ & $\leq0.13$\\
6 & 8.97--10.74 & $\leq0.064$ & $\leq0.065$\\
7 & 9.47--10.55 & $\leq0.032$ & $\leq0.032$\\
8 & 9.73--10.27 & $\leq0.024$ & $\leq0.016$\\
\hline
\end{tabular}
\end{table}

\subsection{Boundary conditions} 
\label{subsection:boundary conditions}

We implemented modified outflow boundary conditions at both the inner and outer radial boundaries. At the inner boundary, instead of copying the gas density and pressure from the innermost active zone to the ghost zones, we extended these quantities using the same power-law profile applied during gas initialization. For gas velocity, the azimuthal component follows a Keplerian profile, while the radial and polar components are copied from the innermost active zone, ensuring that no external mass enters the simulation domain. Reflective boundary conditions are applied in the polar ($\theta$) direction, and periodic boundary conditions are used in the azimuthal direction. We adopt reflective boundary conditions in the polar direction primarily to prevent magnetic flux from artificially escaping through the polar boundaries and to maintain numerical stability. Since the primary focus of this work is the disk region near the midplane and far from the polar boundary, this choice is not expected to qualitatively affect the transport behavior studied here.

\subsection{Initial conditions} \label{initial_conditions_subsection}

We perform simulations initialized in hydrostatic equilibrium, with a magnetic field derived from the vector potential to ensure a divergence-free $B$ field (i.e. $\nabla \cdot {\rm \bf B} = 0$) and initially axisymmetric setup in the azimuthal direction.

 The initial conditions of the gas and the magnetic field are the same as in \cite{Hsu24}, and the initial temperature and gas density profiles follow those used in \cite{Hu22}. Specifically, we divide the simulation domain into a cold and dense disk and a hot low-density corona, maintaining a constant aspect ratio of $h/r=0.05$,  where $h$ is the height of the disk scale. The cold and dense disk is confined within two scale heights above and below the midplane, defined as $\pi/2 - \theta_0 <\theta < \pi/2 + \theta_0$, where $\theta_0=\arctan{(2h/r)}$. The gas density and midplane temperature both follow power-law distributions with indices $p=-1.5$ and $q=-1$, respectively:
\begin{eqnarray}
\rho(r,\pi/2) = \rho_0(r/r_0)^{p},  \\ \nonumber 
T(r,\pi/2) = T_0(r/r_0)^{q} 
\label{eq:profile}
\end{eqnarray}
where $r_0=1$~au is the radius of the inner boundary of the computational domain, and $\rho_0=2.667 \times 10^{-10} {\rm g/cm^{-3}}$ and $T_0=570$~K are the density and temperature at $r_0$. To ensure a smooth transition between the cold disk and hot corona, we adopt the following vertical profile for the temperature:  
\begin{equation}
T(r,\theta)=
\begin{cases}
T(r,\pi/2) & \text{if }  |\theta-\pi/2| < \theta_0 \\
T(r,\pi/2)\ {\rm exp}[(|\theta-\pi/2|\\
\ -\theta_0)/\theta_0 \times\ln(160)]
  & \text{if } \theta_0 \leq |\theta-\pi/2| \leq 2\theta_0  \\
160\ T(r,\pi/2); & \text{if } |\theta-\pi/2| > 2\theta_0\\
\end{cases}\label{eq:T}
\end{equation}
We use a quick $\beta_{\rm cool}$ cooling scheme with a cooling timescale of only $10^{-10}$ of the local orbital period, so the temperature profile is effectively fixed over time.
The vertical density profile is generated based on hydrostatic equilibrium, i.e.,$v_r=v_\theta=0$. 

The initial magnetic field is computed from the magnetic vector potential used in \cite{Zanni07}:
\begin{align}
 & B_r (r, \theta) = \frac{1}{r^2 {\rm sin}\theta} \frac{\partial A_\phi}{\partial\theta}, \label{eq:Br} \\
 & B_\theta (r, \theta) = -\frac{1}{r\ {\rm sin}\theta} \frac{\partial A_\phi}{\partial r}, \label{eq:Btheta} \\
 & B_\phi = 0, \label{eq:Bphi}
\end{align}
with
\begin{align}
 & A_\phi (r, \theta) = \frac{4}{3}r_0^2 B_{\rm p,0} \left( \frac{r {\rm sin}\theta}{r_0}\right)^{\frac{3}{4}} \frac{1}{(1 + 2{\rm cot^2 \theta})^{5/8}}
\end{align}
where $B_{\rm p,0}$ sets the scale for the poloidal field strength. The magnetic field setup is the same as that of \cite{Bai17}, \cite{Suriano18}, and \cite{Hu22}. In all our simulations, $B_{\rm p,0}$ is set by plasma-$\beta = 10^3$ in the mid-plane. 

%\subsubsection{Initial setup of 3D simulations} \label{3D_initial_conditions_subsubsection}
We set the initial total dust mass density $\rho_{\rm d}$ to $1\%$ of the initial gas mass density $\rho_{\rm g}$ and set the initial dust velocity equal to that of the gas. As in \cite{Hsu26}, we use 5 dust species of size $a=0.1$, $1/\sqrt{10}$, 1, $\sqrt{10}$ mm, and 1~cm, following a power-law dust distribution with an index of -3.5. Approximately $45\%$ of the dust mass resides in the 1~cm bin, yielding an initial 1~cm pebble-to-gas mass ratio of $\epsilon_{{\rm p},0}\sim 0.45\%$. In the relatively dense regions within four gas scale heights of the midplane, we computed the dust stopping time based on the dust size $a$:
\begin{align}
 &\tau_{\rm s} = \frac{\rho_{\rm m}}{\rho_{\rm g}} \frac{a}{v_{\rm th}}
\end{align}
where $\rho_{\rm m}$ is the grain material density, taken to be 3~${\rm g/cm^{3}}$ here and in the chemistry network (\S \ref{subsec: chemistry and non-ideal}). $v_{\rm th}$ is the thermal speed of the gas. In the low-density regions above and below four gas scale heights from the midplane, we fix the Stokes number ${\rm St}=\tau_{\rm s}\Omega_{\rm K}$ at $0.01$ to make the simulation run more stably. This assumption does not significantly affect our conclusions since they will be based primarily on grains that settle well below four gas scale heights.  
%=\sqrt{8k_{\rm b}T/\pi \mu m_{\rm H}} 

\subsection{Protoplanet} \label{subsec: planet}

The protoplanet is placed on a fixed circular orbit at $r=$10~au in the disc mid-plane \footnote{We note that the substructures can be affected by planetary migration \citep[e.g.,][]{Kanagawa20} and orbital eccentricity \citep[e.g.,][]{Salcedo23,Pichierri24}, which are not included in our simulation.}.
The combined gravitational potential from the star and planet is given by 
\begin{align}
    & \Phi = - GM_{\star} \left[ \frac{1}{r} + \frac{M_{\rm p}/M_{\star}}{\sqrt{|{\bf r} - {\bf r}_{\rm p}|^2+r_{\rm s}^2}}\right] \label{eq: gravity_soften}
\end{align}
where $M_{\star}$ and $M_{\rm p}$ are the masses of the central star and planet, respectively, ${\bf r}_{\rm p}$ is the displacement vector from the star to the planet, and $r_{\rm s}$ is the softening radius to avoid the gravitational potential approaching infinity near the protoplanet. We set $M_{\star}=1M_\odot$ and 
$r_{\rm s}=0.01$~au. 
%, which is approximately twice the smallest cell size in the planet's Hill sphere. 
We ignore the indirect potential term due to the offset between the central star and the center of gravity of the star–planet system, as it has a negligible impact on the perturbation structure \citep{Miranda19,Hu25}. 

The Hill radius of a planet is defined as \citep{Armitage24}
\begin{align}
 & R_{\rm H} = \left(\frac{M_{\rm p}}{3M_{\star}} \right)^{\frac{1}{3}}R_{\rm p}
\end{align}
where $R_{\rm p}=10$~au is the distance of the planet from the star, and $R_{\rm H} \sim $ 0.149~au in our model with the 0.01 Jupiter mass ($\sim$ 3$M_{\oplus}$) of the protoplanet, which is significantly smaller than the gas scale height at the planet's orbit $H_{\rm p} =$ 0.5~au.  The Bondi radius is given by 
\begin{equation}
R_{\rm B}=\frac{GM_{\rm p}}{c_{\rm s,p}^2},
\end{equation}
where $c_{\rm s,p}$ is the circumstellar midplane sound speed at the planet's orbit. We find $R_{\rm B}\simeq0.038$ au, which yields a dimensionless mass  $m\equiv M_{\rm p}/M_{\rm th}=R_{\rm B}/H_{\rm p}\simeq0.076$. The protoplanet is therefore significantly subthermal. 

\subsection{Chemical network and the non-ideal MHD coefficients} 
\label{subsec: chemistry and non-ideal}

The chemical network and the non-ideal MHD coefficients used in this work are the same as in \cite{Hsu24} (see their Sec. 3 and Appendix A for details). Briefly, we used a simplified chemical network following \cite{Umebayashi90}, \cite{Nishi91}, and \cite{grassi19}, which include the element H, He, C, O, and Mg, an MRN-type \citep{Mathis77}  power-law grain size distribution (with $a_{\rm min} = 0.5$ $\mu$m and $a_{\rm max} = 25$ $\mu$m, and a power-law index of $-3.5$), and gas-phase, gas-grain, and grain-grain reactions. Once the charge abundances are computed from the chemical network, they are stored in a lookup table and referenced during simulations to calculate the Ohmic and ambipolar diffusivities for each computational cell using the standard formulae (see, e.g., Eq.~[18]-[24] of \citealt{Hsu24}).  

%To mimic the expected better magnetic coupling (and thus reduced diffusivity) due to higher ionization levels expected in the lower density regions near the disk surface and in the disk wind, 
To mimic the enhanced magnetic coupling (and thus reduced diffusivity) expected from increased ionization by stellar UV and X-ray irradiation in the lower-density regions near the disk surface and in the disk wind, we follow \cite{Suriano18} and multiply the computed Ohmic and ambipolar diffusivities by the following $\theta$ dependence: 

\begin{align}
 & f(\theta) = 
\begin{cases} 
{\rm exp} \left(-\frac{{\rm cos}^2(\theta + \theta_0)}{2(h/r)^2} \right) & \mbox{if} \quad  \theta < \frac{\pi}{2} -  \theta_0 \\
1 & \mbox{if} \quad \frac{\pi}{2} - \theta_0 \leq \theta   \leq \frac{\pi}{2} + \theta_0 \\
{\rm exp} \left(-\frac{{\rm cos}^2(\theta - \theta_0)}{2(h/r)^2} \right) & \mbox{if} \quad \theta > \frac{\pi}{2} + \theta_0 
\end{cases} .
\label{eq:theta_depend_non_ideal} 
\end{align}
In our setup, ambipolar diffusion dominates over Ohmic dissipation throughout most of the simulation domain, especially in the outer disk and in the low-density disk atmosphere. Near the dense inner-disk midplane, the two diffusivities become more comparable, although ambipolar diffusion generally remains similar to or larger than Ohmic dissipation. Angular momentum transport in the simulations is mediated primarily by large-scale magnetic stresses associated with the magnetized disk wind, especially outside the planet's orbit.

\section{SIMULATION RESULTs: OVERVIEW} \label{sec:results}

We first review the large-scale gas and dust distributions that provide
material to the planet's vicinity, then examine the local concentration
of pebbles within the Hill sphere. This overview establishes the context
for the analysis of the pebble circumplanetary disk (CPD) in Section~\ref{sec:dust in the Hill sphere}, including its dynamical support and its mass supply.

\subsection{Global Gas and Dust Evolution}
\label{subsec: Global}

\begin{figure*}
   \centering
   \includegraphics[width=\linewidth]{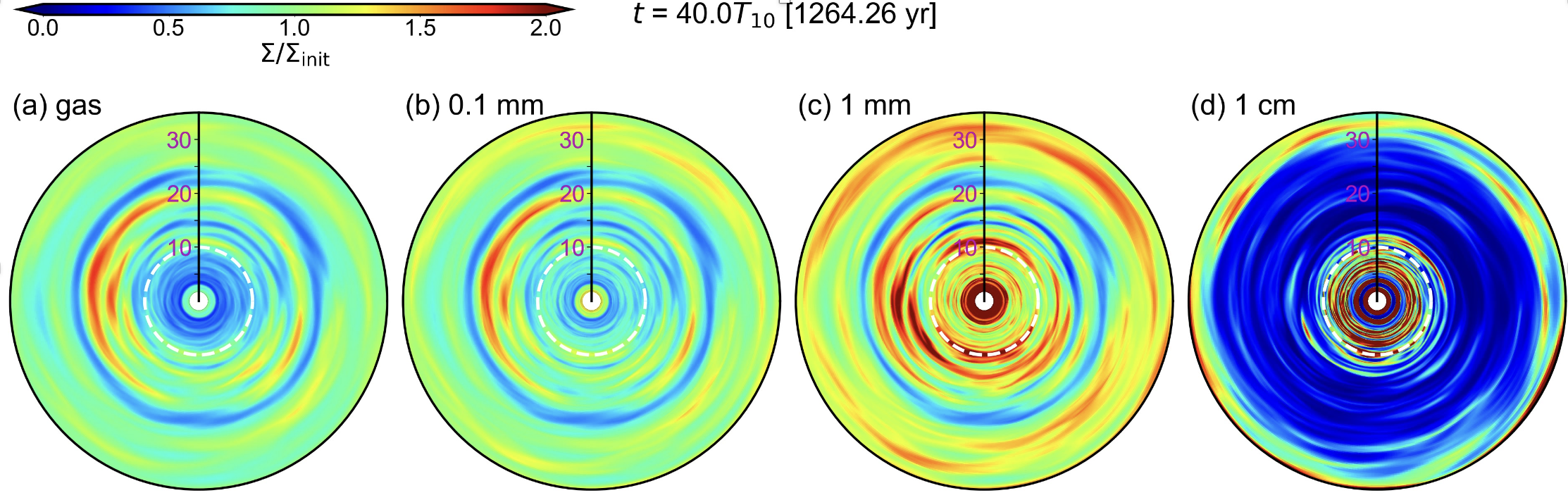}
   \caption{Face-on gas and dust distributions at $t=40T_{10}$ ($\simeq1264$~yr). 
   Panels (a)--(d) show face-on surface densities of gas and grains of size 0.1~mm, 1~mm and 1~cm, respectively, normalized to their
   initial values. The white dashed circles mark the planet's orbit at $10$~au. The planet is located at an azimuthal angle of $\phi=\pi$ (to the left of the center).  An animated version of this figure showing the time evolution of these distributions is available at: \url{https://figshare.com/s/ac0cfa381a31caaa9c9e?file=71275324}.
   }
   \label{fig: M001_5dust_faceonview}
\end{figure*}

\begin{figure*}
   \centering
   \includegraphics[width=\linewidth]{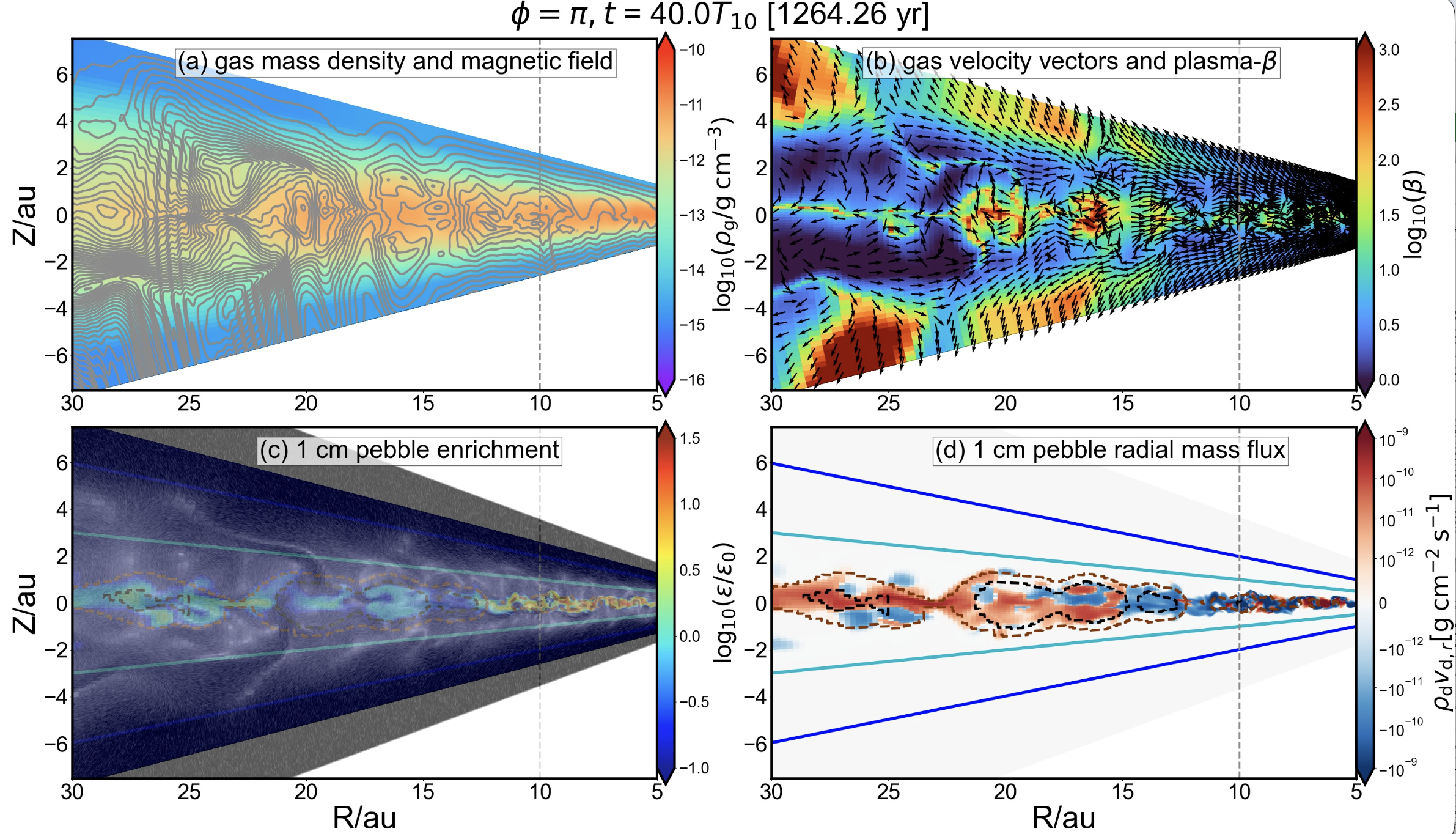}
   \caption{Global structure on the $\phi=\pi$ meridional plane through the planet at $t=40 T_{10}$, with the star to the right.
   Panels show (a) gas density with magnetic field lines; (b) plasma $\beta$ with gas velocity vectors;
   (c) the 1~cm dust-to-gas density ratio relative to its initial value,
   $\log_{10}(\epsilon/\epsilon_0)$, with dust flow visualized using
   line integral convolution (LIC); and (d) the signed radial dust mass
   flux per unit area, $\rho_{\rm d}v_{{\rm d},r}$, for 1~cm grains.
   In (d), blue denotes inward transport and red outward transport. In (c) and
   (d), cyan and blue curves mark two and four gas scale heights,
   respectively, while brown and black dashed contours mark gas
   densities of 0.6 and 1.0 times their initial values, respectively.
   The vertical dashed line marks the planet's orbital radius; $z$ is height above the mid-plane. Circumstellar
   radius decreases toward the right, toward the star.
   An animated version showing the evolution of the gas flow and pebble transport is available at \url{https://figshare.com/s/68944942a0ec1291cdd8?file=71277262}.
   }
   \label{fig: global_meridional}
\end{figure*}

The global evolution of the lower-resolution counterpart of this model was presented and discussed in
\citet{Hsu26}. Here we summarize the features most relevant to pebble
CPD formation and evolution in our higher-resolution simulation with additional levels of static refinement around the protoplanet. 
Figs.~\ref{fig: M001_5dust_faceonview} and
\ref{fig: global_meridional} show representative face-on and meridional
views at $t=40 T_{10}$, where $T_{10}$ is the planet's orbital period at 10~au; their associated movies show the evolution leading to these distributions.

The gas exhibits multiple rings and gaps characteristic of the
magnetized, wind-launching disc
(Fig.~\ref{fig: M001_5dust_faceonview}a).
The dust response depends strongly on grain size: the 0.1~mm grains
broadly follow the gas substructure, whereas the 1~mm and especially
the 1~cm grains show stronger radial redistribution and concentration
(Fig.~\ref{fig: M001_5dust_faceonview}b--d).
The associated movie (see the caption for a link) shows pebbles
migrating across multiple rings and gaps before accumulating at
smaller radii. Thus these substructures regulate the pebble supply without acting
as absolute barriers to inward transport, consistent with
\citet{Hsu26}. The movie also shows an inner region of highly enhanced
1~cm pebble surface density expanding outward with time towards the planet's
orbital neighborhood during much of the evolution, while the region outside the planet's orbit becomes increasingly more depleted. 

The meridional views reveal the vertical structure underlying this
transport. The magnetic field is pinched towards the disc mid-plane
(Fig.~\ref{fig: global_meridional}a; particularly clear
in the early movie frames), and the narrow high-$\beta$ layer in
panel (b) traces the current sheet where the
horizontal magnetic field reverses sign \citep{Hsu26}. This layer is
associated with a narrow mid-plane gas accretion stream at early times.
The 1~cm grains settle into a much thinner layer than the gas and
become concentrated near this stream
(panel c),
allowing radial drift and gas advection to transport them inward
efficiently. The dust enrichment map in panel (c) measures changes in
the local dust-to-gas ratio relative to its initial value. 

The radial flux map of 1~cm pebble (and its associated movie) shows that
this transport is spatially and temporally structured (Fig.~\ref{fig: global_meridional}d).
Initially, inward transport is concentrated in a comparatively coherent
layer near the mid-plane. Later, the layer becomes distorted, with
neighboring regions of inward and outward transport.
The snapshot at $40T_{10}$ illustrates this patchiness
near the orbital radius of the protoplanet, together with mixed inward and
outward transport at larger radii. The movie shows that the gas flow
and the vertical distribution of the pebble-bearing layer also vary, changing the environment from which pebbles approach the protoplanet.
The increasingly distorted pebble layer and the
expanding inner surface-density enhancement suggest a possible
connection between meridional circulation and radial pebble
accumulation. By displacing pebbles from the rapidly accreting
mid-plane stream, circulation may alter their radial transport
and promote local accumulation.
We return to the possible connection between this evolving
circumstellar reservoir and enhanced vertical feeding of the
CPD during Phase~III in
Section~\ref{subsec: CPD flow evolution}.
%\textcolor{red}{[Re-examine the radial extent and timing of the surface-density enhancement and the evolving circulation when higher-cadence movies become available.]}
These variations set the global context for the local pebble supply near the protoplanet.

\subsection{Gas and Dust Evolution in the Hill Sphere}

\begin{figure*}
   \centering
   \includegraphics[width=\linewidth]{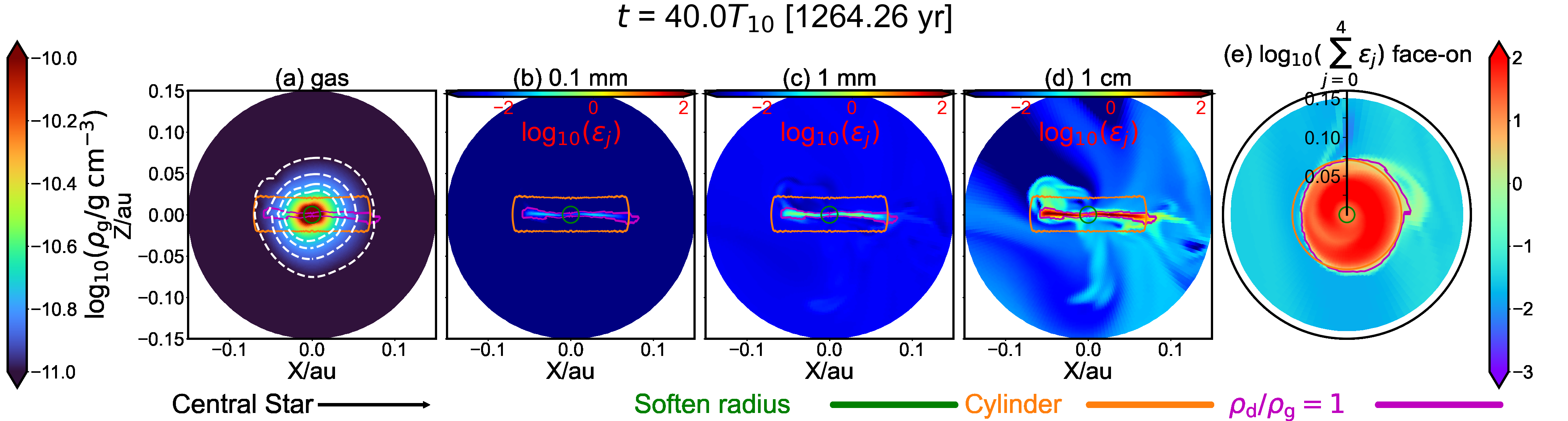}
   \caption{
   Gas and dust distributions within the protoplanet's Hill sphere at
   $t=40T_{10}$, with the star to the right. Panels (a)--(d) show meridional
   slices through the protoplanet for the gas, 0.1~mm, 1~mm and 1~cm grains,
   respectively; panel (e) shows the 1~cm grains in the disc mid-plane.
   The gas map shows gas density, with the isodensity contours highlighting its approximately spherical distribution. The dust maps in (b)--(d) show the logarithm of the dust-to-gas ratio $\epsilon_j$ for each dust species $j$, and the map in (e) shows the total dust-to-gas ratio.  
   Green circles mark the gravitational softening radius.
The orange line marks the analysis cylinder used in
   Section~\ref{subsec: CPD mass evolution} in meridional cross-section for (a)-(d) and mid-plane projection for (e). The magenta contour marks the boundary of the central 
   dust-dominated region, where the local total dust-to-gas ratio
   $\epsilon=\rho_{\rm d}/\rho_{\rm g}=1$. 
   An animated version is available at
\url{https://figshare.com/s/ffdccb885fdb5f5c0876?file=71277340}. 
   }
   \label{fig: Hill-density}
\end{figure*}

We now examine the gas and dust within the protoplanet's Hill sphere,
$R_{\rm H}\simeq0.149$~au.
Fig.~\ref{fig: Hill-density} compares the local distributions at
$t=40T_{10}$, with meridional views of the gas and three representative grain sizes (0.1~mm, 1~mm, and 1~cm) and a complementary mid-plane view of the 1~cm grains.

The gas forms a centrally concentrated, approximately spherical envelope around the protoplanet (panel a), whereas the dust is concentrated towards the
mid-plane (panels b--d). A thin dust layer is visible even for the
0.1~mm grains; its geometry therefore differs substantially from that
of the gas despite the stronger aerodynamic coupling of these grains.
The most pronounced concentration occurs for the 1~cm pebbles, which
form a dense, flattened central structure with a local pebble-to-gas density ratio $\epsilon_{\rm p}$ up to four orders of magnitude higher than its initial value ($\epsilon_{{\rm p},0}$; see Fig.~\ref{fig: CPD Hill phases} below), pushing the total dust-to-gas ratio $\epsilon$ beyond unity by more than one order of magnitude. Its broad extent in the mid-plane view (panel e), compared with its small vertical thickness in panel (d), identifies a disc-like morphology. We refer to this central dust-dominated structure around the protoplanet as a pebble CPD and examine its dynamical support, formation, and evolution in
Section~\ref{sec:dust in the Hill sphere}.

To distinguish the dust-dominated pebble CPD from its surroundings, we use the local total dust-to-gas ratio
$\epsilon=\rho_{\rm d}/\rho_{\rm g}\geq1$ to delineate the central structure for the mass analysis, with $\rho_{\rm d}$ including all simulated dust species. This threshold provides an operational boundary;
its dynamics are examined through force analysis
in Section~\ref{subsec: Force Balance}.
It is marked by a magenta contour in each panel.
Also marked in orange is the fixed analysis cylinder
that approximately encloses the CPD and is used to measure horizontal and vertical
mass exchange in Section~\ref{subsec: CPD mass evolution}. 

The animated version of the figure (see the caption for a link) shows that the gas and dust structures within the Hill sphere evolve substantially over time. The gas distribution remains relatively steady and spherical close to the protoplanet, but becomes more variable and irregular in the outer part of the Hill sphere. The central 1~cm pebble concentration develops around the protoplanet during early evolution
and becomes more prominent at later times, while remaining embedded in changing pebble-rich 
filaments and asymmetric extensions (see panel [d] and its animated version). The ambient dust filaments sometimes bend away from the mid-plane and develop transient concentrations at elevated heights,
while the central pebble CPD around the protoplanet remains identifiable in both the mid-plane and meridional views despite changes in its shape and outer extensions. 
The distributions of 0.1~mm and 1~mm also vary, but their central concentrations are much weaker than those of the 1~cm pebbles. Thus, the pebble CPD is a persistent, cm-sized-grain-dominated flattened structure in a dynamic environment. 

\section{Dynamical Support, Formation and Evolution of the Pebble CPD}
\label{sec:dust in the Hill sphere}

The thin, dust-dominated structure identified in
Section~\ref{sec:results} motivates the questions: what supports the
pebbles against rapid gravitational collapse toward the protoplanet, and how does the structure form initially and grow and evolve at later times? We first
examine its dynamical support at the representative time $t=40T_{10}$ (\S~\ref{subsec: Force Balance}). We then follow the dust mass evolution of the pebble CPD, and seek to understand it using the mass flux through a flattened cylinder enclosing the CPD (\S~\ref{subsec: CPD mass evolution}). We then examine how the mass flux connects to the CPD's larger-scale environment (\S~\ref{subsec: CPD flow evolution}). 

\subsection{Dynamical Support of Pebble CPD}
\label{subsec: Force Balance}

Because our simulation is carried out in a frame corotating with the protoplanet at angular velocity
$\boldsymbol{\Omega}_{\rm p}=\Omega_{\rm p}\hat{z}$, the gas and dust
momentum equations contain the Coriolis acceleration,
$-2\boldsymbol{\Omega}_{\rm p}\times\mathbf{V}_{\rm g,d}$, and the
rotating-frame centrifugal acceleration,
$-\boldsymbol{\Omega}_{\rm p}\times
(\boldsymbol{\Omega}_{\rm p}\times\mathbf r)
=\Omega_{\rm p}^2R\hat{R}$, where $R=r\sin\theta$ is the star-centred cylindrical radius. The gravitational potential $\Phi$
in Eq.~\ref{eq: gravity_soften} contains the stellar and softened
planetary contributions. We combine these gravitational terms with
the rotating-frame centrifugal acceleration to define an \textit{effective gravity}:
\begin{align}
 \mathbf g_{\rm eff}
 &= -\boldsymbol\nabla\Phi+\Omega_{\rm p}^2R\hat{R}
 \nonumber\\
 &= -\frac{GM_\star}{r^2}\hat r
 -\frac{GM_{\rm p}(\mathbf r-\mathbf R_{\rm p})}
 {\left(|\mathbf r-\mathbf R_{\rm p}|^2+r_{\rm s}^2\right)^{3/2}}
 +\Omega_{\rm p}^2R\hat R .
 \label{eq: eff_gravity}
\end{align}
Here $\mathbf R_{\rm p}$ is the protoplanet's position relative to the star,
and $r_{\rm s}=0.01$~au is the gravitational softening radius.
The Coriolis acceleration is treated separately. The centrifugal term
associated with rotation around the protoplanet is also kept separate from
the rotating-frame centrifugal acceleration already included in
$\mathbf g_{\rm eff}$. Magnetic forces have been found to be small
within the CPD, so we neglect them in force analysis.

\begin{figure}
   \centering
   \includegraphics[width=\linewidth]{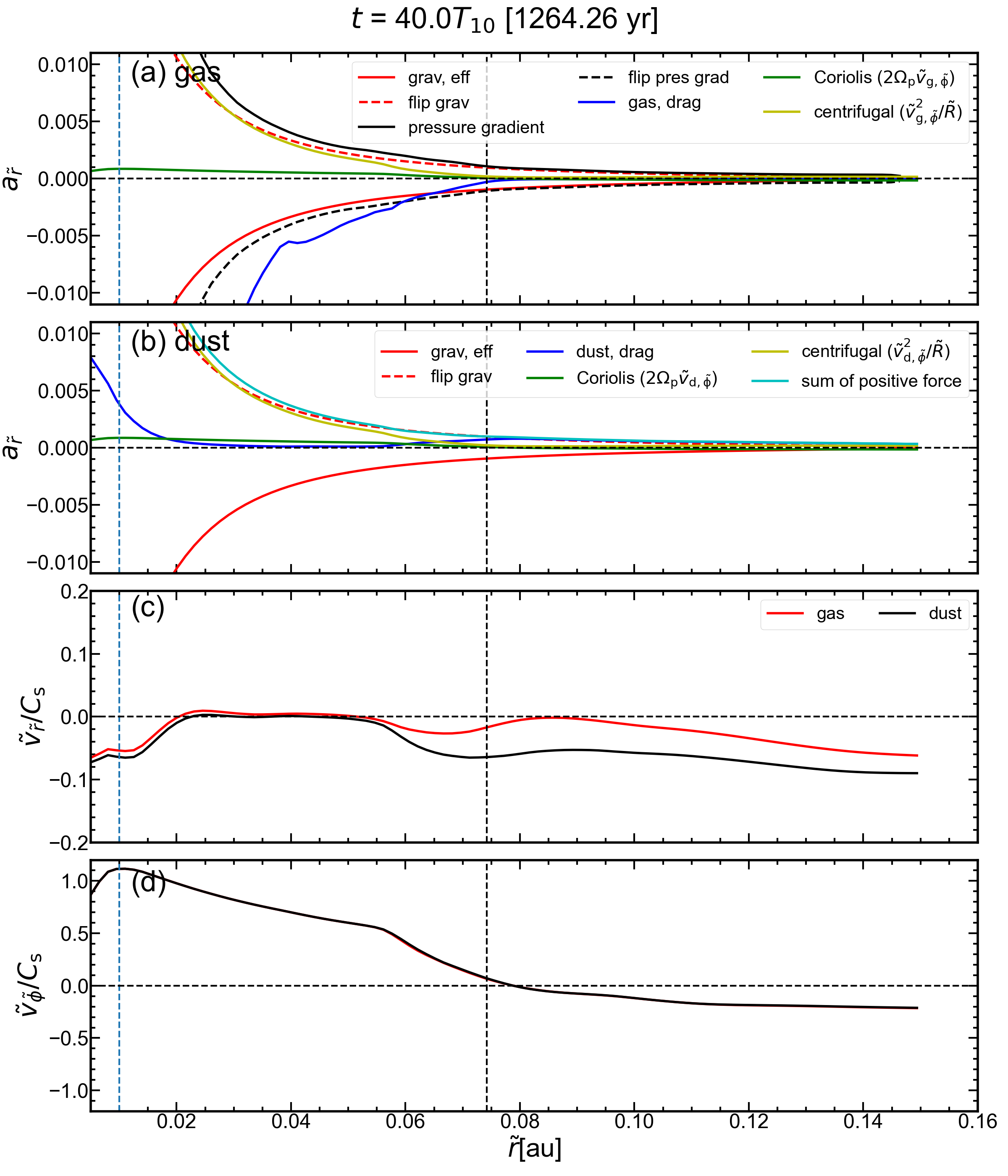}
   \caption{
   The azimuthally ($\tilde\phi$-)averaged radial profiles of the force terms on the midplane at
   $t=40T_{10}$ for gas (panel a) and 1~cm dust (panel b),
   expressed as accelerations in cgs units. The vertical dashed blue line marks
   the softening radius, $r_{\rm s}=0.01$~au, and the vertical dashed
   black line marks the pebble CPD outer boundary (where $\epsilon=1$). Dashed red
   and black curves show the sign-reversed effective gravity and
   pressure acceleration, respectively, for comparison. The cyan curve in (b) is the plotted sum
   of positive contributions. Panel (c) shows the radial speeds of the dust and gas toward the protoplanet in units of the sound speed $(C_s)$. Panel (d) shows that the gas and 1~cm pebbles rotate at essentially the same speed, so the two lines overlap. An animated version is available at \url{https://figshare.com/s/33bcd22dca2141ff8e11?file=71277670}. 
   }
   \label{fig: force_balance}
\end{figure}

Fig.~\ref{fig: force_balance} shows the azimuthally ($\tilde\phi$-)averaged radial
profiles of each force term on the midplane at $t=40\ T_{10}$ for gas and 1~cm
dust. For dust grains (panel [b]), there is no pressure gradient to balance
gravity. Instead, the centrifugal term from rotation around the protoplanet ($\tilde{v}_{d,\phi}^2/\tilde{R}$, where $\tilde{R}$ is the distance from the protoplanet and the tilde denotes a quantity relative to the protoplanet), Coriolis acceleration, and aerodynamic drag contribute to the radial dynamics. Near the outer boundary of the CPD (marked by the black vertical dashed line) and beyond, outward gas drag on the 1~cm pebbles (blue curve in panel~[b]) from their inward drift relative to the gas (panel [c]) dominates the centrifugal (yellow) and Coriolis (green) terms, both of which are small because the 1~cm grains rotate relatively slowly there (see panel [d]). 

As the rotation speed of the 1~cm pebbles $\tilde{v}_{{\rm d},\phi}$ increases towards the protoplanet (panel [d]), both their Coriolis and centrifugal accelerations increase, with the latter increasing faster than the former because it increases with $\tilde{v}_{{\rm d},\phi}^2$ rather than linearly and with decreasing radius. It quickly exceeds the outward drag over the inner part of the CPD, where it provides the main support against the protoplanet's gravity; in the outer part of the CPD, both drag and the centrifugal term contribute significantly to the support. The pebble CPD thus has a primarily rotation-supported inner disk and a drag-assisted outer disk. The velocity profiles provide complementary evidence:
Over most of the inner CPD outside the softening radius, the azimuthally averaged radial velocities of both pebbles and gas are close to zero (panel [c]), whereas their azimuthal
velocities are much higher (panel [d]). Rotation therefore strongly dominates the mean radial motion in this region.

For the gas (panel [a]), the thermal pressure gradient joins the force balance. Outside the CPD (beyond the black vertical dashed line), it balances effective gravity, as expected for a predominantly thermally supported atmosphere, with minor contributions from the centrifugal and Coriolis terms, where gas rotation is relatively slow, as for the 1~cm grains. Inside the pebble CPD, the dust dominates the gas, entraining the latter into near co-rotation (panel [d]), with a corresponding centrifugal acceleration comparable to effective gravity (compare the yellow and red dashed curves in panel [a]). This near-corotation is consistent with dust feedback
reducing the relative azimuthal motion and associated drag torque,
which can suppress inward pebble drift and help maintain the small
mean radial velocities in the inner CPD. 
With the gravity approximately balanced by the centrifugal acceleration, one may expect no need for a pressure gradient. However, the pebbles still drift inward relative to the gas (panel [c]), generating a relatively large inward drag acceleration because the force acts on a relatively small amount of gas. The drag term is comparable to, but somewhat larger than, gravity. It is approximately balanced by the pressure gradient (compare the black dashed and blue curves in panel [a]). 

Outside the pebble CPD, the gas centrifugal acceleration becomes negligible because it is no longer forced to rotate rapidly by dust. The drag acceleration is also small because the gas mass is much higher than the dust mass. These leave the pressure gradient to balance gravity, consistent with the approximately spherical envelope shown in Fig.~\ref{fig: Hill-density}. There is therefore shear between the faster-rotating gas inside the pebble CPD and the more slowly rotating gas outside it. Even inside the CPD, the gas rotates predominantly subsonically outside the softening radius (panel [d]), even though its centrifugal acceleration is comparable to the local gravity. As a result, any rotationally induced flattening of the gas distribution remains minimal, because the CPD occupies a small fraction of the volume and the pressure gradient remains dynamically important. 

\subsection{Mass Evolution and Feeding of the Pebble CPD}
\label{subsec: CPD mass evolution}

\begin{figure}
   \centering
   \includegraphics[width=\linewidth]{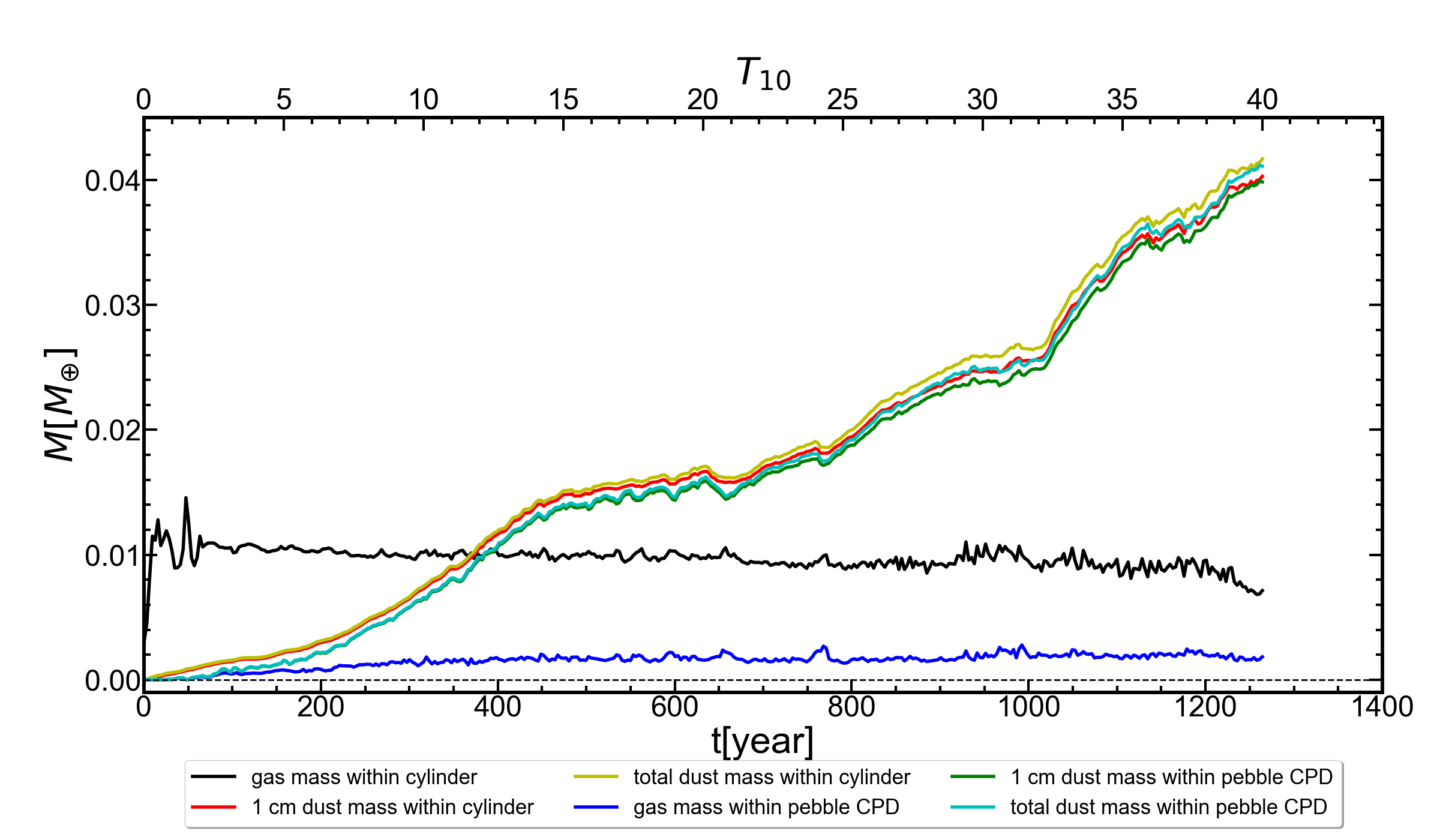}
   \caption{
   Time evolution of the gas and dust masses within the pebble CPD and
   the fixed analysis cylinder enclosing it. Gas masses are shown in
   black (cylinder) and blue (CPD), 1~cm pebble masses in red
   (cylinder) and green (CPD), and total dust masses in yellow
   (cylinder) and cyan (CPD). The CPD is delineated using the local
   total dust-to-gas ratio $\epsilon\geq1$. Masses are in $M_\oplus$;
   the lower axis gives time in years, and the upper axis gives
   $t/T_{10}$.
   }
   \label{fig: CPD cylinder mass}
\end{figure}

\begin{figure}
   \centering
   \includegraphics[width=\linewidth]{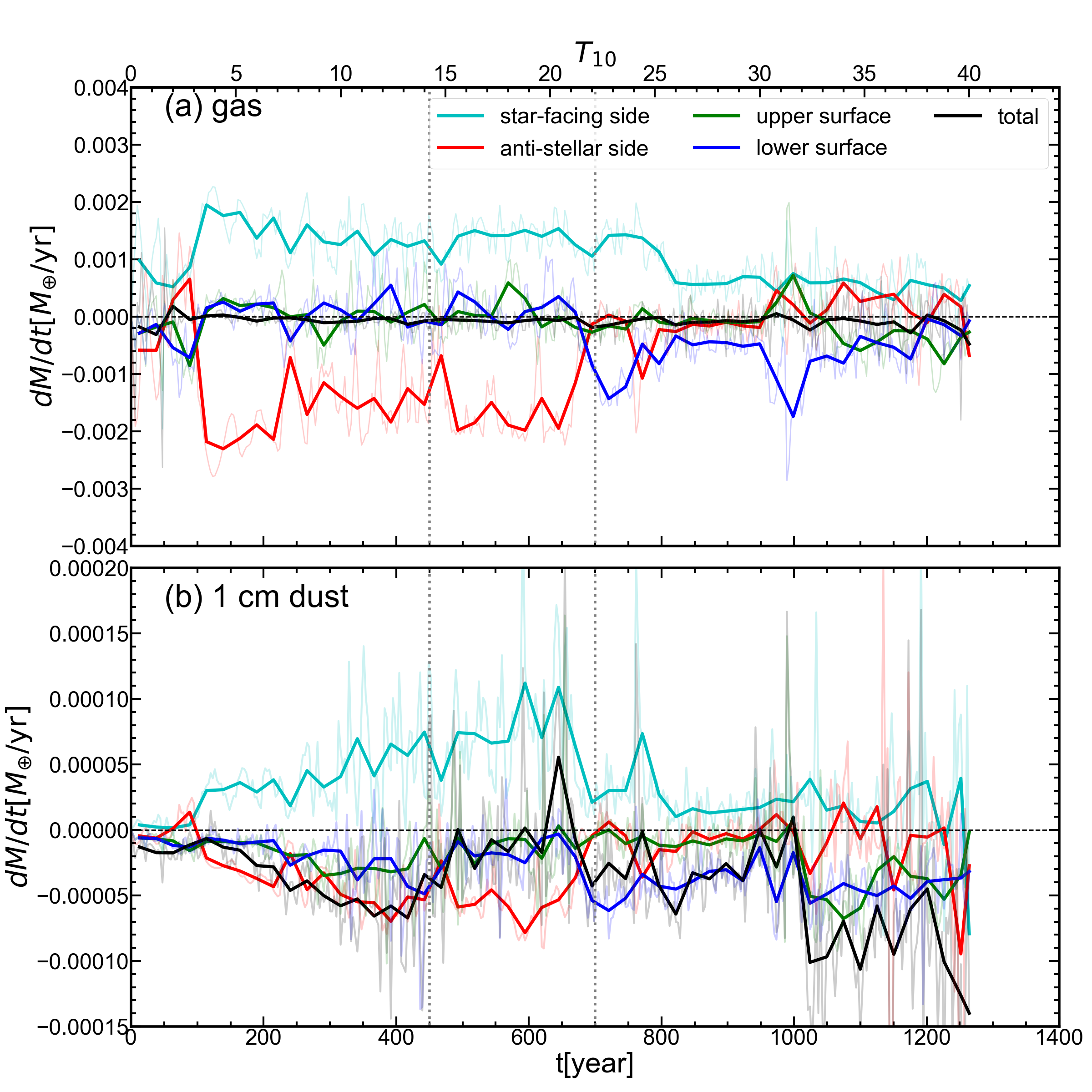}
   \caption{
   Signed mass fluxes through the fixed analysis-cylinder boundary
   for gas (a) and 1~cm pebbles (b). Faint curves show the
   high-cadence flux histories, while thicker curves show the
   approximately 25-yr-averaged contributions. Cyan and red denote 
   the star-facing and anti-stellar curved side halves, respectively;
   green and blue denote the upper and lower surfaces. Black shows
   the net flux. Negative values represent net feeding and positive
   values net loss. The dotted vertical lines at 450 and 700~yr mark the approximate boundaries between Phases~I, II, and III. 
   }
   \label{fig: cylinder face flux}
\end{figure}

Having established the pebble CPD as a primarily rotation-supported, drag-assisted flattened structure, we now follow its mass evolution and examine how it gains and loses mass through a fixed control volume. We use a flattened protoplanet-centered cylinder marked in Fig.~\ref{fig: Hill-density}, with radius
0.07~au and half-height 0.02~au, which approximately encloses the CPD. We separately measure the masses
within this cylinder and within the dust-dominated CPD defined in
Section~3.2 by the local total dust-to-gas ratio
$\epsilon=\rho_{\rm d}/\rho_{\rm g}\geq1$.

Fig.~\ref{fig: CPD cylinder mass} compares the gas, 1~cm pebble, and
total dust masses in the cylinder and the CPD. The 1~cm pebbles dominate the
solid mass in both regions. Their mass rises rapidly
during early evolution, reaching  $\sim 1.5\times 10^{-2}M_\oplus$ within the first several hundred years. The rise slows for a few hundred years, before resuming after about $700$~years. The pebble mass within the CPD closely follows that within the cylinder throughout. The gas provides a useful contrast. After an initial adjustment, its
mass within the cylinder remains $\sim 10^{-2}M_\oplus$, without the general increase seen in the pebbles. The gas mass inside the CPD is only a fraction of that in the cylinder, typically of order $2\times10^{-3}M_\oplus$, which is not surprising since the CPD is significantly thinner than the cylinder. It is much lower than the pebble mass in the same region (compare the green and blue curves). Indeed, the ratio of these two curves indicates that the pebble-to-gas mass ratio averaged over the CPD, $\epsilon_{\rm p}$, generally increases with time, reaching values around 20 at the end of the simulation at $40~T_{10}$. 

The close correspondence between the pebble masses inside the CPD and the cylinder motivates using the cylinder boundary fluxes to investigate CPD feeding and
loss approximately. We divide the boundary into four surfaces. The
curved lateral boundary is split into a star-facing half, nearer the
star, and an anti-stellar half, facing larger circumstellar radii.
These halves measure the integrated horizontal exchange over the full
 height of the cylinder. The upper and lower faces measure vertical exchange. Each contribution is a signed net flux through that surface and can
include simultaneous inflow and outflow in different areas.

In Fig.~\ref{fig: cylinder face flux}, negative flux denotes feeding and positive flux denotes loss. The high-cadence curves (thinner and fainter curves in the background) resolve
strong variability, including brief episodes of net outward pebble transport (panel [b]), while the 25-year averages (thicker curves) show the balance between feeding and loss more clearly. The gas initially exhibits strong
anti-stellar-side entry and star-facing-side exit. These opposing
horizontal contributions remain prominent throughout much of the first
700~yr. They largely cancel, leaving a small net flux compared with
 the individual contributions, consistent with the relatively steady
gas inventory. The vertical gas fluxes vary in sign, with outward
transport during many early intervals. After approximately 700~yr, the anti-stellar-side inflow weakens markedly, and the star-facing outflow also decreases, but more gradually, while inward transport through the lower surface becomes prominent. The gas exchange therefore acquires a stronger vertical contribution at late times.

For pebbles, anti-stellar-side feeding and star-facing loss also provide substantial opposing contributions during the early and
middle evolution. In contrast to the gas, the upper and lower surfaces generally provide net feeding. During early intervals with
outward vertical gas flux, this opposing transport is consistent with pebbles settling towards the CPD relative to the gas. At later
times, vertically inward fluxes for both the gas and the dust through the lower surface instead suggest a contribution from gas-advection-assisted vertical pebble delivery.

The mass histories and surface fluxes suggest three broad phases of the CPD evolution, with approximate boundaries at 450 and 700~yr or $14.2T_{10}$ and $22.1T_{10}$. 

\textit{Phase I: assembly and rapid growth
($0\lesssim t\lesssim450$~yr).} Anti-stellar-side pebble feeding, complemented by inflow through the upper and lower surfaces, generally exceeds star-facing loss. The predominantly negative net pebble (feeding) flux accompanies rapid growth
to $\sim 1.5\times10^{-2}M_\oplus$ within the CPD. 
Gas exchange is dominated by opposing horizontal inflow and outflow; its vertical transport is frequently outward while pebbles enter vertically. Horizontal supply and vertical settling therefore both contribute to early pebble accumulation.

\textit{Phase II: slower accumulation with enhanced horizontal exit ($450\lesssim t\lesssim700$~yr).}
The pebble mass increases more slowly, with fluctuations around
its overall upward trend. Anti-stellar-side supply persists, but star-facing exit reaches larger values and offsets more of the
feeding. The high-cadence curves resolve strong outward-flux bursts, including episodes in which the total pebble flux becomes positive. These episodes interrupt accumulation. 
%Vertical pebble delivery remains active but is weaker during parts of this interval than in the strong feeding episodes towards the end of Phase~I. Together, these changes reduce the net accumulation rate. 

\textit{Phase III: renewed growth with enhanced vertical feeding ($700\lesssim t\lesssim1264$~yr).} Both the anti-stellar-side pebble inflow and the opposing star-facing outflow weaken compared with the preceding phases. Vertical feeding
becomes increasingly more important, particularly from the lower surface.  The high-cadence curves show that persistent vertical delivery coexists with short episodes of net loss. 

These flux patterns motivate an examination of the surrounding flow structures in the next subsection. 

\subsection{Flow Structure During the CPD Formation and Evolution}
\label{subsec: CPD flow evolution}

\begin{figure*}
 \centering
 \includegraphics[width=\linewidth]{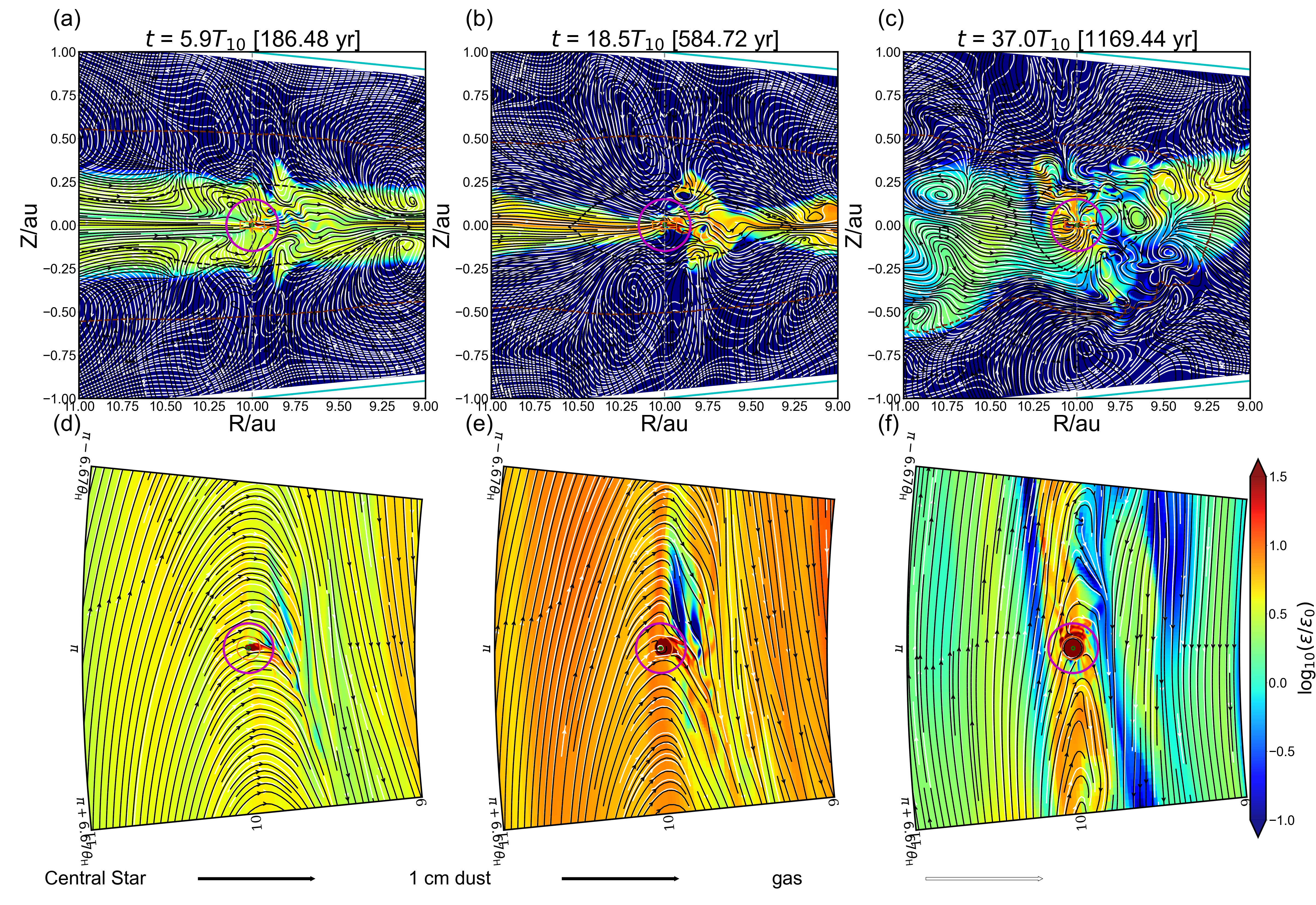}
 \caption{
 Intermediate-scale environment of the pebble CPD at $t=5.9$, $18.5$
 and $37T_{10}$ (approximately 186, 585 and 1169~yr), representing
 Phases~I, II and III from left to right. Panels (a)--(c) are
 meridional views through the star--protoplanet plane; panels
 (d)--(f) are the corresponding mid-plane views. The star is towards
 the right. Colour shows the 1~cm dust-to-gas density ratio relative
 to its initial value, $\log_{10}(\epsilon/\epsilon_0)$, over the
 range $-1$ to $1.5$. Black and white streamlines trace the
 instantaneous dust and gas velocity fields, respectively. The
 magenta circle marks the Hill sphere, and the orange rectangle in
 the meridional panels marks the analysis cylinder. The brown and
 black dashed gas-density reference contours and cyan scale-height
 curves are as in Fig.~\ref{fig: global_meridional}.
 An animated version is available at {\url{https://figshare.com/s/863007b3fb609633990e?file=71734318}}.
 }
 \label{fig: CPD intermediate phases}
\end{figure*}

\begin{figure*}
 \centering
 \includegraphics[width=\linewidth]{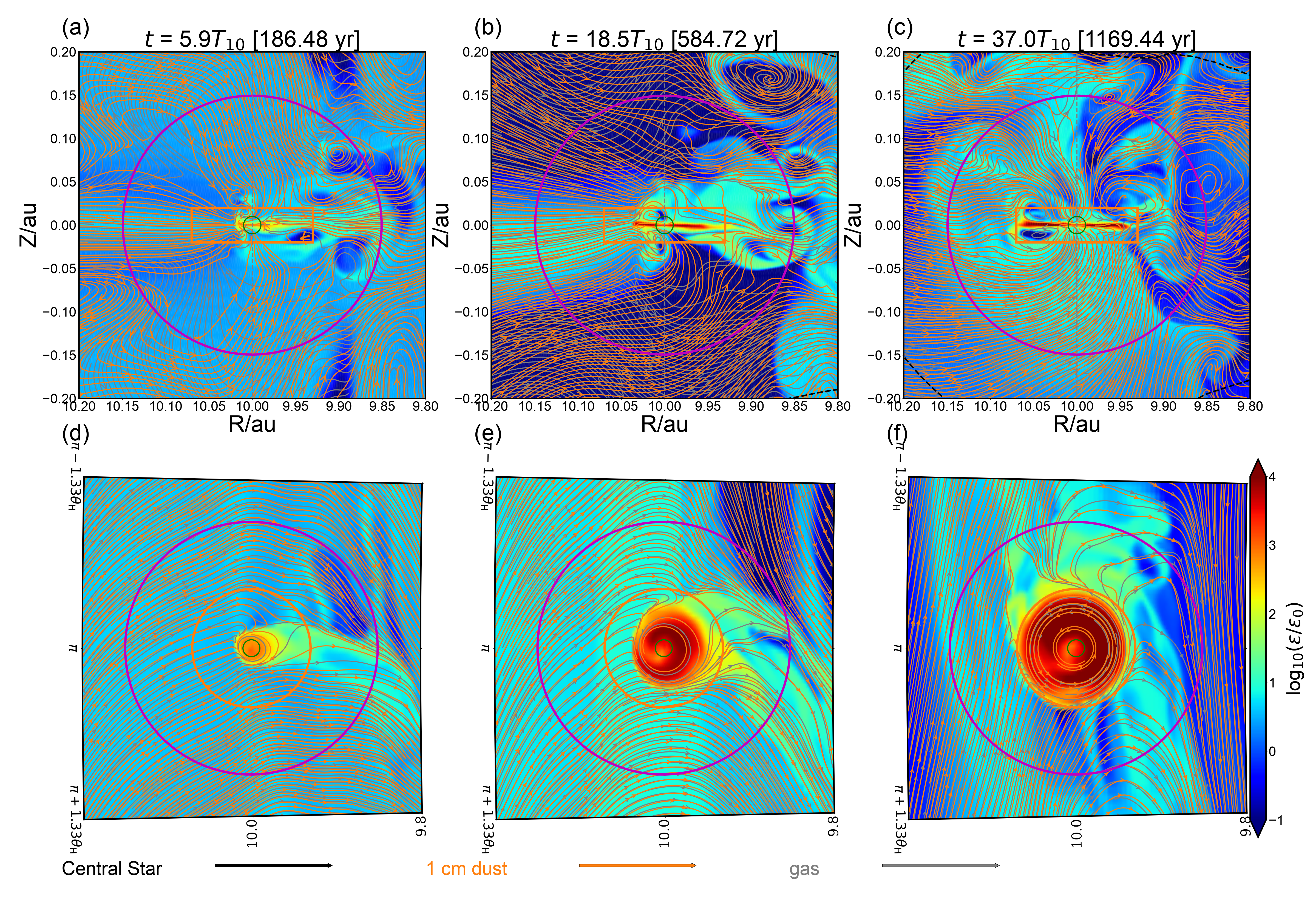}
 \caption{
 Hill-scale structure of the pebble CPD at the same representative epochs and in
 the same panel arrangement as Fig.~\ref{fig: CPD intermediate phases}.
 The star is towards the right. Colour shows $\log_{10}(\epsilon/\epsilon_0)$ for 1~cm pebbles over the range
 $-1$ to $4$. Orange and fainter grey streamlines trace the instantaneous dust and gas velocity fields, respectively. The
 magenta circle marks the Hill sphere, the green circle marks the
 gravitational softening radius, $r_{\rm s}=0.01$~au, and the orange
 rectangle in panels (a)--(c) and circle in panels (d)--(f) the analysis cylinder. 
 An animated version is available at {\url{https://figshare.com/s/ad21fb29169b168066fd?file=71734321}}.
 }
 \label{fig: CPD Hill phases}
\end{figure*}

We illustrate the three phases using representative epochs
$t=5.9$, $18.5$, and $37T_{10}$ (approximately 186, 585, and 1169~yr). 
Fig.~\ref{fig: CPD intermediate phases} shows the intermediate-scale
environment, while Fig.~\ref{fig: CPD Hill phases} resolves the
corresponding structure near and inside the Hill sphere. Each column
pairs the same epoch across both spatial scales and viewing planes. The associated movies show the evolution between snapshots, with
the global context provided by Section~\ref{subsec: Global}. The plotted streamlines
trace instantaneous velocity fields and illustrate approximate delivery
and exit routes; the analysis cylinder surface-integrated fluxes in Fig.~\ref{fig: cylinder face flux} establish their net contributions.

\textit{Phase I: assembly and early growth.}
As discussed in Section~\ref{subsec: Global}, the 1~cm pebbles settle towards the circumstellar disk mid-plane. Gas accretion driven by magnetic
braking carries them inward, together with aerodynamic drift relative to the gas. At $5.9 T_{10}$, the intermediate-scale meridional
view (Fig.~\ref{fig: CPD intermediate phases}a) shows a relatively thick pebble-bearing layer extending above and below the planet's Hill sphere. The mid-plane view (panel~d) shows the circumstellar accretion flow entering the Hill sphere, where the protoplanet can modify it. 

The modification is clearly shown in the Hill-scale mid-plane view (Fig.~\ref{fig: CPD Hill phases}d), where dust streamlines spiral around the protoplanet in the prograde direction. This geometry is consistent with the capture of part of the inward-moving pebble supply into the growing CPD. It
is similar in spirit to the prograde inspiral of captured particles illustrated in the semi-analytic work of \citet[his Fig.~12, panel~E]{Ormel2013}, although circumstellar radial accretion is stronger in our simulation.

\begin{figure}
 \centering
 \includegraphics[width=\linewidth]{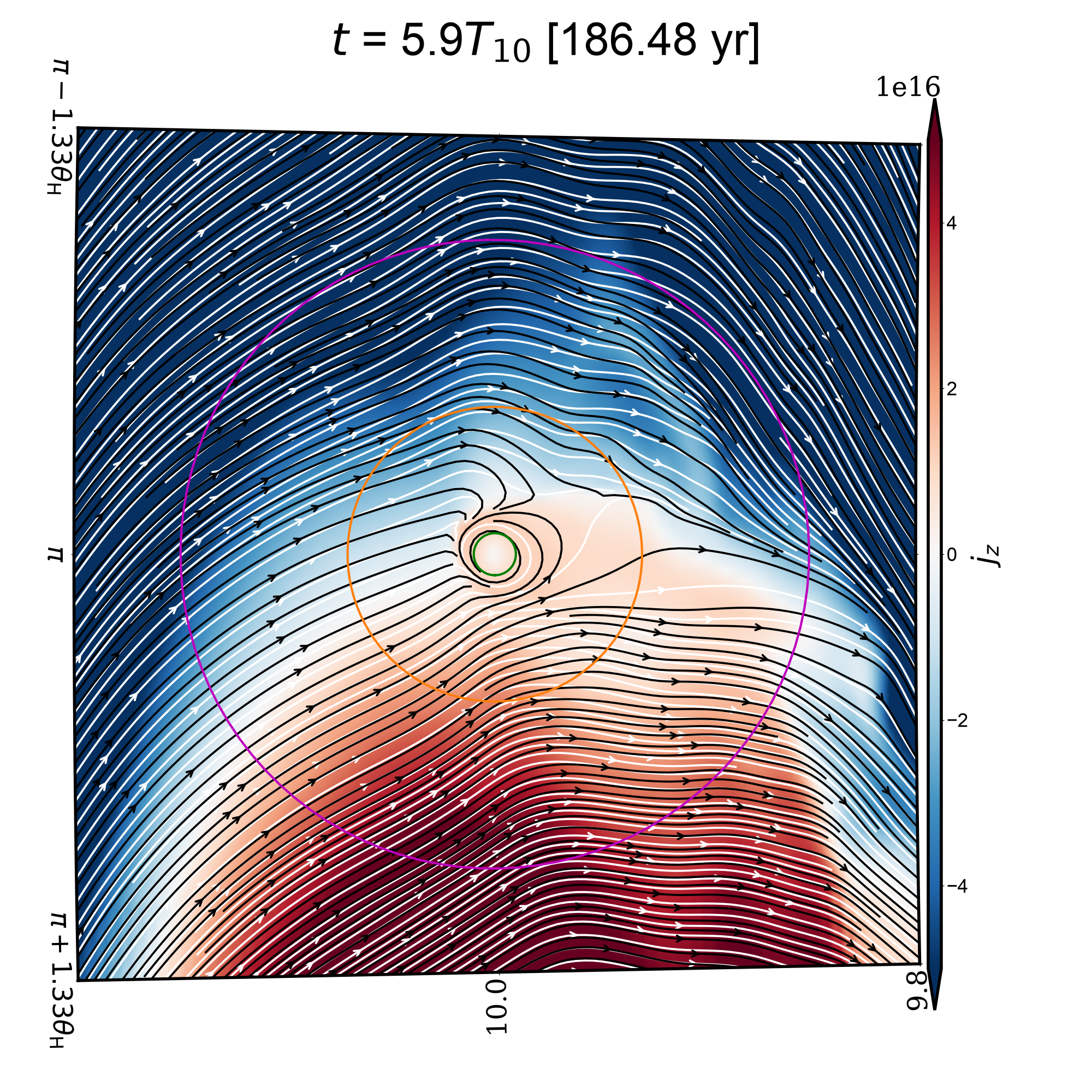}
 \caption{
 Planet-centred specific angular momentum $J_z$ of the 1~cm pebbles
 in the mid-plane at $t=5.9~T_{10}$ (approximately 186~yr), the same
 Phase~I epoch used in Figs.~\ref{fig: CPD intermediate phases}
 and \ref{fig: CPD Hill phases}. The star is towards the right.
 Colour shows $J_z$ in $\mathrm{cm^2\,s^{-1}}$, computed using a non-rotating frame centered on the planet; positive
 values denote prograde motion. Black and white streamlines trace 
 the instantaneous pebble and gas velocity fields in the corotating
 frame, respectively. The magenta, orange and green circles mark the Hill radius, analysis-cylinder radius and gravitational softening
 radius, respectively.
 An animated version will be available at {\url{https://figshare.com/s/43ea370c0a6dcb9357ce?file=71734324}}.
 }
 \label{fig: CPD pebble angular momentum}
\end{figure}

The planet-centered specific angular momentum of the 1~cm pebbles provides a complementary view of this early capture geometry (Fig.~\ref{fig: CPD pebble angular momentum}). We compute its
vertical component as
\begin{equation}
J_z=\widetilde R\left(\widetilde v_{{\rm d},\phi}
+\Omega_{\rm p}\widetilde R\right),
\end{equation}
where $\widetilde R$ is the cylindrical distance from the vertical axis through the planet, and $\widetilde v_{{\rm d},\phi}$ is the azimuthal velocity in the frame centered on the planet and corotating with its orbit. The term $\Omega_{\rm p}\widetilde R^2$ converts the angular momentum from the corotating frame to a planet-centered frame that does not co-rotate with the planet's orbit. Positive $J_z$ denotes prograde motion.

At the representative time $5.9~T_{10}$, a broad region surrounding the protoplanet in the plotted
mid-plane view carries positive $J_z$, particularly below (or upstream of) the planet, where it reaches several $10^{16}\,\mathrm{cm^2\,s^{-1}}$. For comparison, rotational support requires $J_z$ of 
$\sim 10^{16}\,\mathrm{cm^2\,s^{-1}}$ near the gravitational
softening radius (green circle) and
$\sim 3\times10^{16}\,\mathrm{cm^2\,s^{-1}}$ near the analysis-cylinder boundary (orange circle), which approximately encloses the well-formed pebble CPD at later times (see Fig.~\ref{fig: Hill-density}).
Thus, parts of the apparent feeding region carry sufficient specific angular momentum, in principle, to support orbits on CPD scales.  

The Hill-scale meridional view
(Fig.~\ref{fig: CPD Hill phases}a) shows pebbles approaching the mid-plane 
pebble-rich CPD around the protoplanet from elevated heights. This behavior is consistent with the Phase~I pebble flux pattern: pebble
feeding through the upper and lower surfaces complements the horizontal feeding through the anti-stellar
side, with net exit through the star-facing side. 

\textit{Phase II: slower accumulation and enhanced horizontal loss.}
At the representative time $18.5T_{10}$, the intermediate-scale meridional view
(Fig.~\ref{fig: CPD intermediate phases}b) still shows a vertically extended 
pebble-bearing layer outside the protoplanet's orbit, although it does not extend above and below the Hill sphere, unlike in the earlier frame at $5.9~T_{10}$. The thinner layer is more concentrated near the mid-plane, as shown in the mid-plane view (panel~e; compare its color with that of panel~d), which likewise shows dust streams entering the Hill region. The slower accumulation in this phase therefore does not reflect a lack of horizontal pebble supply from larger radii. The Hill-scale meridional view (Fig.~\ref{fig: CPD Hill phases}b)
shows a thin, well-formed CPD with a pronounced star-facing extension, which appears to be associated with vigorous meridional circulation, as indicated by the curvy dust streamlines and multiple thin dust-rich filaments; the circulation is shown more clearly in the animated version of the figure (see the caption for a link). Nevertheless, the circulation does not appear to disrupt the pebble CPD, which remains clearly identifiable in both the midplane and meridional view. 

\textit{Phase III: vertical delivery assisted by gas advection.}
At the representative time $37T_{10}$, the intermediate-scale meridional view
(Fig.~\ref{fig: CPD intermediate phases}c) shows the pebble-bearing
layer outside the planet's orbit becomes less concentrated than at the intermediate time $18.5~T_{10}$ because stronger meridional circulation displaces the layer away from the midplane and makes it thicker.  The Hill sphere is filled with elevated pebble-rich filaments (see the orange filaments inside the purple circle in panel~c), which appear to be part of a larger ring-like pebble concentration near the planet's orbit, as shown in the mid-plane view (panel~f, the orange strip around 10~au). This ring-like concentration may be associated with the outward expansion of the dust-enhanced inner region and the increasingly warped pebble layer seen in the global movies (Section~\ref{subsec: Global}). Together, these developments suggest
that the evolving circumstellar flow provides a highly structured and variable 
pebble reservoir near the planet's orbit, from which elevated
filaments can supply the CPD. 
%\textcolor{red}{[Check this proposed connection and its timing relative to the renewed CPD growth when higher-cadence movies become available.]}

The Hill-scale meridional view (Fig.~\ref{fig: CPD Hill phases}c)
shows that the vertically elevated pebble filaments enter the analysis cylinder from the upper and, especially, the lower surface together with the gas, feeding the CPD. Because the dust-to-gas mass ratio is well below unity outside the CPD, advection by the relatively dense, converging gas flow towards the CPD helps deliver pebbles to the CPD in this epoch, along with gravitational settling. However, unlike pebbles, the gas does not accumulate near the planet because pressure buildup limits its accumulation. The Hill scale middle plane view (Fig.~\ref{fig: CPD Hill phases}f) shows a well-formed pebble CPD rotating around the protoplanet with local pebble enhancement up to $10^4$ of their initial value. However, the enhancement is not uniform across the entire CPD, likely reflecting the patchy, dynamic nature of its feeding from a highly inhomogeneous and variable environment.

\section{Discussion}
\label{sec:discussion}

\subsection{Pebble CPD formation within a low-mass protoplanet's envelope}
\label{subsec:pebble_CPD_formation}

To put the formation and persistence of pebble CPD in a broader context, it is useful to compare our results with calculations of dust transport through planetary envelopes. A particularly relevant comparison is provided by
\citet{Kuwahara2026}, who follow dust transport through
convectively stable envelopes of embedded protoplanets
using three-dimensional multifluid hydrodynamic simulations. Their dimensionless planetary mass is
$m\equiv R_{\rm B}/H_{\rm p}=M_{\rm p}/M_{\rm th}=0.1$, close to
$m\simeq0.076$ in our model. Their ratios
$R_{\rm H}/H_{\rm p}\simeq0.32$ and $R_{\rm H}/R_{\rm B}\simeq3.2$
are also similar to our values of approximately 0.29 and 3.9,
respectively. Both calculations therefore lie
in the subthermal regime with $R_{\rm B}<R_{\rm H}<H_{\rm p}$.
Their results share some features, but also differ in important ways.

In the three-dimensional calculation of \citet{Kuwahara2026},
the larger particles settle towards the midplane and produce a
flattened dust distribution within the gas envelope (their Fig.~7c),
similar in morphology to our pebble CPD. However, their midplane
dust-to-gas ratio is much lower than ours, remaining at or below
its initial value of $\sim1\%$ and decreasing near the protoplanet
(their Fig.~5b). Rapid radial infall limits the
particles' residence time.  
Indeed, their largest particles approach the free-fall speed in
the inner envelope, while rotation remains dynamically subdominant
over the spatially resolved region (their Fig.~4).
The increasing azimuthal speed towards the
protoplanet nevertheless leaves open the possibility of a smaller
rotationally supported structure inside their fiducial inner
boundary at $0.1R_{\rm B}$. In contrast, our simulation produces a spatially resolved, dust-dominated CPD whose azimuthal speed exceeds its radial infall speed.

This difference may result from a combination of the dust loading of the incoming material
and dust feedback. As shown in
Fig.~\ref{fig: CPD intermediate phases} and its animation,
vertical settling in the global circumstellar disk has already enhanced the concentration of 1~cm pebbles
outside the protoplanet's orbit by up to an order of magnitude
before they enter the Hill sphere. Further concentration inside
the Hill sphere pushes the local dust-to-gas ratio above unity. Dust feedback then promotes near-corotation
of the gas and pebbles 
(see Fig.~\ref{fig: force_balance}d), reducing their relative azimuthal motion and
the associated drag torque. This can help pebbles remain in
rotationally supported orbits, allowing further accumulation
and strengthening the collective resistance to angular-momentum loss through gas drag.
This interpretation is consistent with the pioneering work of
\citet{Johansen2010}, who found
that higher initial dust-to-gas mass ratios
favoured the formation of a rotationally supported dust CPD
when dust feedback was included (see their Fig.~2),
even around a very low-mass object of dimensionless mass
$10^{-6}$. They also showed that accretion from the prograde particle disk delivers angular momentum to the central body, potentially spinning it up. Our pebble CPD may similarly influence the protoplanet’s spin, although quantifying this effect requires higher-resolution calculations that follow the transfer of mass and angular momentum to the planet.

\citet{Krapp2022} also investigated the dust distribution inside
the Hill sphere of a more massive subthermal protoplanet,
with a dimensionless planet mass of about 0.6
(compared to our 0.076). They also find a flattened dust
distribution inside the Hill sphere (see their Fig.~2), but its vertical thickness appears to be set primarily by the settling--diffusion equilibrium established outside the Hill sphere. Because they focus on the effects of the dust distribution on opacity, they
do not establish whether the concentrated dust forms a rotationally supported, dust-dominated CPD. Their work nevertheless highlights the potential importance of such a structure for the opacity distribution in the protoplanet's envelope.

The discussion above points to high dust loading and dust feedback as key ingredients
for pebble CPD formation around subthermal protoplanets. Settling of the relatively large grains traced by millimeter continuum emission is observed in many circumstellar disks \citep[e.g.,][]{Villenave2025}. 
The enhanced dust concentration outside the
Hill sphere, coupled with further settling inside the approximately spherical gaseous envelope,
favours the formation of a flattened pebble distribution.
Whether this distribution becomes 
rotationally supported depends on the angular momentum of the captured pebbles
and its subsequent evolution.
The incoming angular momentum is set by 
the dust and gas flow structure in the circumstellar disk,
including radial accretion towards the star. In our simulation,
the relatively fast midplane accretion, particularly during Phase~I, may influence the angular
momentum delivered by the captured pebbles and hence the size of the resulting CPD. Once the CPD becomes dust-dominated,
feedback may slow subsequent angular-momentum loss and help it persist.
Under other dust and gas flow conditions, the pebbles' incoming angular momentum and its removal
rate may differ, producing a different CPD size or preventing
its formation. Nevertheless, the large separation between
the Hill radius and the much smaller physical radius of the protoplanet (by three orders of magnitude) leaves room for at least compact pebble CPDs, provided that captured pebbles retain sufficient angular momentum to orbit outside the protoplanet. Such structures may remain unresolved in global disk simulations.

\subsection{Implications for Pebble Delivery and Satellite Formation}

The pebble CPD introduces an intermediate reservoir between
capture from the circumstellar disc and delivery to the
protoplanet. As shown in
Section~\ref{sec:dust in the Hill sphere}, captured pebbles
can accumulate in a persistent circumplanetary structure while
continuing to exchange material with their surroundings.
Entry into the Hill sphere therefore need not imply prompt
accretion onto the protoplanet. The fraction of this reservoir that ultimately reaches the planet, or remains available for other outcomes, remains to be determined.

One possible outcome is the formation of satellite building
blocks. As reviewed by \citet{Nimmo2026}, major uncertainties
in satellite formation include how solids are supplied to
circumplanetary space, how they avoid rapid inward drift,
and whether they can form satellitesimal seeds.
Our results show that a dust-dominated, rotating pebble reservoir
can develop around a low-mass protoplanet even when its gaseous
envelope remains approximately spherical. Thus, potential satellite-building material can accumulate without a rotationally supported gaseous CPD on the same spatial scales.
Dust feedback may help retain this material by reducing its
relative azimuthal motion with respect to the gas
(Section~\ref{subsec:pebble_CPD_formation}).
The resulting solid-rich reservoir may be relevant
to gas-starved satellite-formation models
\citep{Canup2002}, whose lower gas surface densities and gradual
supply can permit cooler conditions, slower
satellite growth, and reduced inward Type~I migration compared
to the classical minimum-mass subnebula
\citep{Nimmo2026}. Although our pebble CPD is gas-poor relative
to its solid content, whether it provides these conditions
remains to be established. Nevertheless, it could in principle supply material for satellitesimal formation (potentially through processes such as streaming instability in the CPD where the dust-to-gas mass ratio exceeds unity) or the growth of existing seeds through pebble accretion, potentially allowing satellite
assembly to begin before the protoplanet completes its growth.

\section{Conclusions}
\label{sec:conclusion}

We used global three-dimensional non-ideal MHD simulations to follow the structure and evolution of gas
and dust from a magnetic wind-launching circumstellar disk to a three-Earth-mass protoplanet orbiting
at 10~au. By resolving the flow within the Hill
sphere and including aerodynamic feedback from dust on gas, we examined how a circumplanetary pebble disk (pebble CPD) forms, what supports it, and how it
exchanges material with its surroundings. Our main conclusions are as follows.
\smallskip

\noindent (1) The subthermal protoplanet hosts a thin, prograde pebble CPD without opening a deep gap in the circumstellar gas disk. In the pebble CPD, the local 1~cm pebble-to-gas density ratio reaches up to $10^4$ times its initial value, making it dust-dominated, with an average pebble-to-gas mass ratio reaching approximately 20. The radial force profiles indicate that rotation provides the main support against the planetary gravity in the inner pebble disk, with aerodynamic drag contributing
farther out. The small azimuthally averaged radial velocities compared with the rotation speeds provide complementary evidence for the dynamical support of the inner CPD. The surrounding gas remains predominantly pressure supported and forms an approximately spherical envelope.

\smallskip

\noindent (2) The pebble CPD assembles rapidly, fed by pebbles drifting inward from larger stellar radii and settling toward it from above and below. Its pebble mass reaches $\sim 10^{-2}\,M_\oplus$ within the
first several hundred years. Accumulation slows during an intermediate phase (approximately 450--700~yr), when enhanced intermittent horizontal losses offset more of the feeding. Growth resumes after about 700~yr, with enhanced vertical delivery assisted by gas advection. The dust-dominated CPD
persists through the end of the simulation, reaching $\sim 4\times 10^{-2}\,M_\oplus$ in about 1200~yr. Dust feedback may help maintain it by bringing the gas into near-corotation with the pebbles, thereby reducing their angular momentum loss through aerodynamic drag.
\smallskip

\noindent (3) The pebble CPD provides a reservoir between capture into the Hill sphere and delivery to the protoplanet, potentially providing material for early satellite formation. However, it remains open how much of this material ultimately reaches the protoplanet or forms satellite precursors.

\section*{Acknowledgements}

We thank Stanley Baronett and Leo Krapp for useful discussions. CYH acknowledges support from the NRAO ALMA Student Observing Support (SOS) and computing resources from UVA research computing (RIVANNA) and NSF ACCESS (AST200032, PHY240192). This work is supported in part by NSF AST-2307199, JWST-GO-02104.002-A, JWST-GO-08872.003-A, and the Virginia Institute of Theoretical Astronomy (VITA). MKL is supported by the National Science and Technology Council (grants 113-2112-M-001-036-, 114-2112-M-001-018-, 113-2124-M-002-003-, and 114-2124-M-002-003-), an Academia Sinica Career Development Award (AS-CDA-110-M06), and an Academia Sinica Grand Challenge Seed Program (AS-GCS-115-M02).

%%%%%%%%%%%%%%%%%%%%%%%%%%%%%%%%%%%%%%%%%%%%%%%%%%
\section*{Data Availability}

The data underlying this article will be shared on reasonable request
to the corresponding author.



\bibliographystyle{mnras}
\bibliography{ref} % if your bibtex file is called example.bib

@ARTICLE{Fung2019,
       author = {{Fung}, Jeffrey and {Zhu}, Zhaohuan and {Chiang}, Eugene},
        title = "{Circumplanetary Disk Dynamics in the Isothermal and Adiabatic Limits}",
      journal = {\apj},
         year = 2019,
        month = dec,
       volume = {887},
       number = {2},
          eid = {152},
        pages = {152},
          doi = {10.3847/1538-4357/ab53da},
archivePrefix = {arXiv},
       eprint = {1909.09655},
 primaryClass = {astro-ph.EP},
       adsurl = {https://ui.adsabs.harvard.edu/abs/2019ApJ...887..152F}
}

@ARTICLE{Bethune2019,
       author = {{B{\'e}thune}, William and {Rafikov}, Roman R.},
        title = "{Envelopes of embedded super-Earths - II. Three-dimensional isothermal simulations}",
      journal = {\mnras},
         year = 2019,
        month = sep,
       volume = {488},
       number = {2},
        pages = {2365-2379},
          doi = {10.1093/mnras/stz1870},
archivePrefix = {arXiv},
       eprint = {1907.02763},
 primaryClass = {astro-ph.EP},
       adsurl = {https://ui.adsabs.harvard.edu/abs/2019MNRAS.488.2365B}
}

@ARTICLE{Villenave2025,
       author = {{Villenave}, Marion and {Rosotti}, Giovanni P. and {Lambrechts}, Michiel and {Ziampras}, Alexandros and {Pinte}, Christophe and {M{\'e}nard}, Fran{\c{c}}ois and {Stapelfeldt}, Karl R. and {Duch{\^e}ne}, Gaspard and {Baylock}, Emily and {Doi}, Kiyoaki},
        title = "{Turbulence in protoplanetary disks: A systematic analysis of dust settling in 33 disks}",
      journal = {\aap},
         year = 2025,
        month = may,
       volume = {697},
          eid = {A64},
        pages = {A64},
          doi = {10.1051/0004-6361/202553822},
archivePrefix = {arXiv},
       eprint = {2503.05872},
 primaryClass = {astro-ph.SR},
       adsurl = {https://ui.adsabs.harvard.edu/abs/2025A&A...697A..64V}
}

@ARTICLE{Nimmo2026,
       author = {{Nimmo}, Francis and {Canup}, Robin and {Fujii}, Yuri and {Izidoro}, Andr{\'e} and {Ji}, Jianghui and {McKinnon}, William and {Mousis}, Olivier and {Ormel}, Chris and {Ogihara}, Masahiro and {Stevenson}, Dave and {Blanc}, Michel},
        title = "{Origin and Evolution of the Galilean Satellites Within the Jovian System}",
      journal = {\ssr},
         year = 2026,
        month = may,
       volume = {222},
       number = {4},
          eid = {46},
        pages = {46},
          doi = {10.1007/s11214-026-01295-6},
       adsurl = {https://ui.adsabs.harvard.edu/abs/2026SSRv..222...46N}
}

@ARTICLE{Krapp2022,
       author = {{Krapp}, Leonardo and {Kratter}, Kaitlin M. and {Youdin}, Andrew N.},
        title = "{The 3D Dust and Opacity Distribution of Protoplanets in Multifluid Global Simulations}",
      journal = {\apj},
         year = 2022,
        month = apr,
       volume = {928},
       number = {2},
          eid = {156},
        pages = {156},
          doi = {10.3847/1538-4357/ac5899},
archivePrefix = {arXiv},
       eprint = {2110.02428},
 primaryClass = {astro-ph.EP},
       adsurl = {https://ui.adsabs.harvard.edu/abs/2022ApJ...928..156K}
}

@ARTICLE{Johansen2010,
       author = {{Johansen}, Anders and {Lacerda}, Pedro},
        title = "{Prograde rotation of protoplanets by accretion of pebbles in a gaseous environment}",
      journal = {\mnras},
         year = 2010,
        month = may,
       volume = {404},
       number = {1},
        pages = {475-485},
          doi = {10.1111/j.1365-2966.2010.16309.x},
archivePrefix = {arXiv},
       eprint = {0910.1524},
 primaryClass = {astro-ph.EP},
       adsurl = {https://ui.adsabs.harvard.edu/abs/2010MNRAS.404..475J}
}

@ARTICLE{Kuwahara2026,
       author = {{Kuwahara}, Ayumu and {Lambrechts}, Michiel},
        title = "{Dust transport in envelopes of disk-embedded planets: I. Convectively stable envelopes}",
      journal = {\aap},
         year = 2026,
        month = oct,
       volume = {714},
          eid = {A117},
        pages = {A117},
          doi = {10.1051/0004-6361/202659284},
       adsurl = {https://ui.adsabs.harvard.edu/abs/2026A&A...714A.117K}
}

@ARTICLE{Ormel2013,
       author = {{Ormel}, C.~W.},
        title = "{The steady-state flow pattern past gravitating bodies}",
      journal = {\mnras},
         year = 2013,
        month = feb,
       volume = {428},
       number = {4},
        pages = {3526-3542},
          doi = {10.1093/mnras/sts289},
archivePrefix = {arXiv},
       eprint = {1210.7501},
 primaryClass = {astro-ph.EP},
       adsurl = {https://ui.adsabs.harvard.edu/abs/2013MNRAS.428.3526O}
}

@ARTICLE{Lambrechts12,
       author = {{Lambrechts}, Michiel and {Johansen}, Anders},
        title = "{Rapid growth of gas-giant cores by pebble accretion}",
      journal = {\aap},
         year = 2012,
        month = aug,
       volume = {544},
          eid = {A32},
        pages = {A32},
          doi = {10.1051/0004-6361/201219127},
archivePrefix = {arXiv},
       eprint = {1205.3030},
 primaryClass = {astro-ph.EP},
       adsurl = {https://ui.adsabs.harvard.edu/abs/2012A&A...544A..32L}
}

@ARTICLE{Hu22,
       author = {{Hu}, Xiao and {Li}, Zhi-Yun and {Zhu}, Zhaohuan and {Yang}, Chao-Chin},
        title = "{Formation of dust rings and gaps in non-ideal MHD discs through meridional gas flows}",
      journal = {\mnras},
         year = 2022,
        month = oct,
       volume = {516},
       number = {2},
        pages = {2006-2022},
          doi = {10.1093/mnras/stac1799},
archivePrefix = {arXiv},
       eprint = {2203.05629},
 primaryClass = {astro-ph.EP},
       adsurl = {https://ui.adsabs.harvard.edu/abs/2022MNRAS.516.2006H}
}

@ARTICLE{Umebayashi90,
       author = {{Umebayashi}, Toyoharu and {Nakano}, Takenori},
        title = "{Magnetic flux loss from interstellar clouds}",
      journal = {\mnras},
         year = 1990,
        month = mar,
       volume = {243},
        pages = {103-113},
          doi = {10.1093/mnras/243.1.103},
       adsurl = {https://ui.adsabs.harvard.edu/abs/1990MNRAS.243..103U}
}

@ARTICLE{Nishi91,
       author = {{Nishi}, Ryoichi and {Nakano}, Takenori and {Umebayashi}, Toyoharu},
        title = "{Magnetic Flux Loss from Interstellar Clouds with Various Grain-Size Distributions}",
      journal = {\apj},
         year = 1991,
        month = feb,
       volume = {368},
        pages = {181},
          doi = {10.1086/169682},
       adsurl = {https://ui.adsabs.harvard.edu/abs/1991ApJ...368..181N}
}

@ARTICLE{grassi19,
       author = {{Grassi}, T. and {Padovani}, M. and {Ramsey}, J.~P. and {Galli}, D. and {Vaytet}, N. and {Ercolano}, B. and {Haugb{\o}lle}, T.},
        title = "{The challenges of modelling microphysics: ambipolar diffusion, chemistry, and cosmic rays in MHD shocks}",
      journal = {\mnras},
         year = 2019,
        month = mar,
       volume = {484},
       number = {1},
        pages = {161-184},
          doi = {10.1093/mnras/sty3519},
archivePrefix = {arXiv},
       eprint = {1901.00504},
 primaryClass = {astro-ph.GA},
       adsurl = {https://ui.adsabs.harvard.edu/abs/2019MNRAS.484..161G}
}

@ARTICLE{Bai17,
       author = {{Bai}, Xue-Ning and {Stone}, James M.},
        title = "{Hall Effect-Mediated Magnetic Flux Transport in Protoplanetary Disks}",
      journal = {\apj},
         year = 2017,
        month = feb,
       volume = {836},
       number = {1},
          eid = {46},
        pages = {46},
          doi = {10.3847/1538-4357/836/1/46},
archivePrefix = {arXiv},
       eprint = {1612.03912},
 primaryClass = {astro-ph.EP},
       adsurl = {https://ui.adsabs.harvard.edu/abs/2017ApJ...836...46B}
}

@ARTICLE{Mathis77,
       author = {{Mathis}, J.~S. and {Rumpl}, W. and {Nordsieck}, K.~H.},
        title = "{The size distribution of interstellar grains.}",
      journal = {\apj},
         year = 1977,
        month = oct,
       volume = {217},
        pages = {425-433},
          doi = {10.1086/155591},
       adsurl = {https://ui.adsabs.harvard.edu/abs/1977ApJ...217..425M}
}

@ARTICLE{Suriano18,
       author = {{Suriano}, Scott S. and {Li}, Zhi-Yun and {Krasnopolsky}, Ruben and {Shang}, Hsien},
        title = "{The formation of rings and gaps in magnetically coupled disc-wind systems: ambipolar diffusion and reconnection}",
      journal = {\mnras},
         year = 2018,
        month = jun,
       volume = {477},
       number = {1},
        pages = {1239-1257},
          doi = {10.1093/mnras/sty717},
archivePrefix = {arXiv},
       eprint = {1712.06217},
 primaryClass = {astro-ph.SR},
       adsurl = {https://ui.adsabs.harvard.edu/abs/2018MNRAS.477.1239S}
}

@ARTICLE{Zanni07,
       author = {{Zanni}, C. and {Ferrari}, A. and {Rosner}, R. and {Bodo}, G. and {Massaglia}, S.},
        title = "{MHD simulations of jet acceleration from Keplerian accretion disks. The effects of disk resistivity}",
      journal = {\aap},
         year = 2007,
        month = jul,
       volume = {469},
       number = {3},
        pages = {811-828},
          doi = {10.1051/0004-6361:20066400},
archivePrefix = {arXiv},
       eprint = {astro-ph/0703064},
 primaryClass = {astro-ph},
       adsurl = {https://ui.adsabs.harvard.edu/abs/2007A&A...469..811Z}
}

@ARTICLE{Cui21,
       author = {{Cui}, Can and {Bai}, Xue-Ning},
        title = "{Global three-dimensional simulations of outer protoplanetary discs with ambipolar diffusion}",
      journal = {\mnras},
         year = 2021,
        month = oct,
       volume = {507},
       number = {1},
        pages = {1106-1126},
          doi = {10.1093/mnras/stab2220},
archivePrefix = {arXiv},
       eprint = {2106.10167},
 primaryClass = {astro-ph.EP},
       adsurl = {https://ui.adsabs.harvard.edu/abs/2021MNRAS.507.1106C}
}

@ARTICLE{stone20,
       author = {{Stone}, James M. and {Tomida}, Kengo and {White}, Christopher J. and {Felker}, Kyle G.},
        title = "{The Athena++ Adaptive Mesh Refinement Framework: Design and Magnetohydrodynamic Solvers}",
      journal = {\apjs},
         year = 2020,
        month = jul,
       volume = {249},
       number = {1},
          eid = {4},
        pages = {4},
          doi = {10.3847/1538-4365/ab929b},
archivePrefix = {arXiv},
       eprint = {2005.06651},
 primaryClass = {astro-ph.IM},
       adsurl = {https://ui.adsabs.harvard.edu/abs/2020ApJS..249....4S}
}

@ARTICLE{Suriano17,
       author = {{Suriano}, Scott S. and {Li}, Zhi-Yun and {Krasnopolsky}, Ruben and {Shang}, Hsien},
        title = "{Rings and gaps produced by variable magnetic disc winds and avalanche accretion streams - I. Axisymmetric resistive MHD simulations}",
      journal = {\mnras},
         year = 2017,
        month = jul,
       volume = {468},
       number = {4},
        pages = {3850-3868},
          doi = {10.1093/mnras/stx735},
archivePrefix = {arXiv},
       eprint = {1702.01565},
 primaryClass = {astro-ph.SR},
       adsurl = {https://ui.adsabs.harvard.edu/abs/2017MNRAS.468.3850S}
}

@ARTICLE{Riols19,
       author = {{Riols}, A. and {Lesur}, G.},
        title = "{Spontaneous ring formation in wind-emitting accretion discs}",
      journal = {\aap},
         year = 2019,
        month = may,
       volume = {625},
          eid = {A108},
        pages = {A108},
          doi = {10.1051/0004-6361/201834813},
archivePrefix = {arXiv},
       eprint = {1904.07910},
 primaryClass = {astro-ph.EP},
       adsurl = {https://ui.adsabs.harvard.edu/abs/2019A&A...625A.108R}
}

@ARTICLE{Huang22,
       author = {{Huang}, Pinghui and {Bai}, Xue-Ning},
        title = "{A Multifluid Dust Module in Athena++: Algorithms and Numerical Tests}",
      journal = {\apjs},
         year = 2022,
        month = sep,
       volume = {262},
       number = {1},
          eid = {11},
        pages = {11},
          doi = {10.3847/1538-4365/ac76cb},
archivePrefix = {arXiv},
       eprint = {2206.01023},
 primaryClass = {astro-ph.EP},
       adsurl = {https://ui.adsabs.harvard.edu/abs/2022ApJS..262...11H}
}

@ARTICLE{Hsu24,
       author = {{Hsu}, Chun-Yen and {Li}, Zhi-Yun and {Tu}, Yisheng and {Hu}, Xiao and {Lin}, Min-Kai},
        title = "{Rossby wave instability and substructure formation in 3D non-ideal MHD wind-launching discs}",
      journal = {\mnras},
         year = 2024,
        month = sep,
       volume = {533},
       number = {3},
        pages = {2980-2996},
          doi = {10.1093/mnras/stae1986},
archivePrefix = {arXiv},
       eprint = {2407.08032},
 primaryClass = {astro-ph.EP},
       adsurl = {https://ui.adsabs.harvard.edu/abs/2024MNRAS.533.2980H}
}

@ARTICLE{Bethune17,
       author = {{B{\'e}thune}, William and {Lesur}, Geoffroy and {Ferreira}, Jonathan},
        title = "{Global simulations of protoplanetary disks with net magnetic flux. I. Non-ideal MHD case}",
      journal = {\aap},
         year = 2017,
        month = apr,
       volume = {600},
          eid = {A75},
        pages = {A75},
          doi = {10.1051/0004-6361/201630056},
archivePrefix = {arXiv},
       eprint = {1612.00883},
 primaryClass = {astro-ph.EP},
       adsurl = {https://ui.adsabs.harvard.edu/abs/2017A&A...600A..75B}
}

@ARTICLE{Huang25,
       author = {{Huang}, Pinghui and {Bai}, Xue-Ning},
        title = "{Dust Clumping in Outer Protoplanetary Disks: The Interplay among Four Instabilities}",
      journal = {\apjl},
         year = 2025,
        month = jun,
       volume = {986},
       number = {1},
          eid = {L13},
        pages = {L13},
          doi = {10.3847/2041-8213/adcebb},
archivePrefix = {arXiv},
       eprint = {2503.11076},
 primaryClass = {astro-ph.EP},
       adsurl = {https://scixplorer.org/abs/2025ApJ...986L..13H}
}

@ARTICLE{Aoyama23,
       author = {{Aoyama}, Yuhiko and {Bai}, Xue-Ning},
        title = "{Three-dimensional Global Simulations of Type-II Planet-Disk Interaction with a Magnetized Disk Wind. I. Magnetic Flux Concentration and Gap Properties}",
      journal = {\apj},
         year = 2023,
        month = mar,
       volume = {946},
       number = {1},
          eid = {5},
        pages = {5},
          doi = {10.3847/1538-4357/acb81f},
archivePrefix = {arXiv},
       eprint = {2302.01514},
 primaryClass = {astro-ph.EP},
       adsurl = {https://ui.adsabs.harvard.edu/abs/2023ApJ...946....5A}
}

@ARTICLE{Miranda19,
       author = {{Miranda}, Ryan and {Rafikov}, Roman R.},
        title = "{On the Planetary Interpretation of Multiple Gaps and Rings in Protoplanetary Disks Seen By ALMA}",
      journal = {\apjl},
         year = 2019,
        month = jun,
       volume = {878},
       number = {1},
          eid = {L9},
        pages = {L9},
          doi = {10.3847/2041-8213/ab22a7},
archivePrefix = {arXiv},
       eprint = {1905.08259},
 primaryClass = {astro-ph.EP},
       adsurl = {https://ui.adsabs.harvard.edu/abs/2019ApJ...878L...9M}
}

@ARTICLE{Hu25,
       author = {{Hu}, Xiao and {Li}, Zhi-Yun and {Bae}, Jaehan and {Zhu}, Zhaohuan},
        title = "{3D gap opening in non-ideal MHD protoplanetary discs: asymmetric accretion, meridional vortices, and observational signatures}",
      journal = {\mnras},
         year = 2025,
        month = jan,
       volume = {536},
       number = {2},
        pages = {1374-1388},
          doi = {10.1093/mnras/stae2681},
archivePrefix = {arXiv},
       eprint = {2403.18292},
 primaryClass = {astro-ph.EP},
       adsurl = {https://ui.adsabs.harvard.edu/abs/2025MNRAS.536.1374H}
}

@ARTICLE{Kanagawa20,
       author = {{Kanagawa}, Kazuhiro D. and {Nomura}, Hideko and {Tsukagoshi}, Takashi and {Muto}, Takayuki and {Kawabe}, Ryohei},
        title = "{Model of a Gap Formed by a Planet with Fast Inward Migration}",
      journal = {\apj},
         year = 2020,
        month = apr,
       volume = {892},
       number = {2},
          eid = {83},
        pages = {83},
          doi = {10.3847/1538-4357/ab781e},
archivePrefix = {arXiv},
       eprint = {2002.08217},
 primaryClass = {astro-ph.EP},
       adsurl = {https://ui.adsabs.harvard.edu/abs/2020ApJ...892...83K}
}

@ARTICLE{Salcedo23,
       author = {{S{\'a}nchez-Salcedo}, F.~J. and {Chametla}, R.~O. and {Chrenko}, O.},
        title = "{Estimating the depth of gaps opened by planets in eccentric orbit}",
      journal = {\mnras},
         year = 2023,
        month = jan,
       volume = {518},
       number = {1},
        pages = {439-455},
          doi = {10.1093/mnras/stac2856},
archivePrefix = {arXiv},
       eprint = {2210.09008},
 primaryClass = {astro-ph.EP},
       adsurl = {https://ui.adsabs.harvard.edu/abs/2023MNRAS.518..439S}
}

@ARTICLE{Pichierri24,
       author = {{Pichierri}, Gabriele and {Bitsch}, Bertram and {Lega}, Elena},
        title = "{A Recipe for Eccentricity and Inclination Damping for Partial-gap Opening Planets in 3D Disks}",
      journal = {\apj},
         year = 2024,
        month = jun,
       volume = {967},
       number = {2},
          eid = {111},
        pages = {111},
          doi = {10.3847/1538-4357/ad3dff},
archivePrefix = {arXiv},
       eprint = {2404.02247},
 primaryClass = {astro-ph.EP},
       adsurl = {https://ui.adsabs.harvard.edu/abs/2024ApJ...967..111P}
}

@ARTICLE{Armitage24,
       author = {{Armitage}, Philip J.},
        title = "{Planet formation theory: an overview}",
      journal = {arXiv e-prints},
         year = 2024,
        month = dec,
          eid = {arXiv:2412.11064},
        pages = {arXiv:2412.11064},
          doi = {10.48550/arXiv.2412.11064},
archivePrefix = {arXiv},
       eprint = {2412.11064},
 primaryClass = {astro-ph.EP},
       adsurl = {https://ui.adsabs.harvard.edu/abs/2024arXiv241211064A}
}

@ARTICLE{Hsu25,
       author = {{Hsu}, Chun-Yen and {Li}, Zhi-Yun and {Tu}, Yisheng and {Hu}, Xiao and {Lin}, Min-Kai},
        title = "{Dust concentration via coupled vertical settling and radial migration in substructured non-ideal MHD discs and early planet formation}",
      journal = {\mnras},
         year = 2025,
        month = jun,
       volume = {540},
       number = {1},
        pages = {1363-1377},
          doi = {10.1093/mnras/staf798},
archivePrefix = {arXiv},
       eprint = {2503.24142},
 primaryClass = {astro-ph.EP},
       adsurl = {https://ui.adsabs.harvard.edu/abs/2025MNRAS.540.1363H}
}

@ARTICLE{Ormel10,
       author = {{Ormel}, C.~W. and {Klahr}, H.~H.},
        title = "{The effect of gas drag on the growth of protoplanets. Analytical expressions for the accretion of small bodies in laminar disks}",
      journal = {\aap},
         year = 2010,
        month = sep,
       volume = {520},
          eid = {A43},
        pages = {A43},
          doi = {10.1051/0004-6361/201014903},
archivePrefix = {arXiv},
       eprint = {1007.0916},
 primaryClass = {astro-ph.EP},
       adsurl = {https://ui.adsabs.harvard.edu/abs/2010A&A...520A..43O}
}

@ARTICLE{Bai2013a,
       author = {{Bai}, Xue-Ning and {Stone}, James M.},
        title = "{Wind-driven Accretion in Protoplanetary Disks. I. Suppression of the Magnetorotational Instability and Launching of the Magnetocentrifugal Wind}",
      journal = {\apj},
         year = 2013, volume = {769}, eid = {76},
          doi = {10.1088/0004-637X/769/1/76},
       eprint = {1301.0318},
       adsurl = {https://ui.adsabs.harvard.edu/abs/2013ApJ...769...76B}
}

@ARTICLE{Lesur2021,
       author = {{Lesur}, G.},
        title = "{Magnetohydrodynamics of protoplanetary discs}",
      journal = {Journal of Plasma Physics},
         year = 2021, volume = {87}, eid = {205870101},
          doi = {10.1017/S0022377820001002},
       eprint = {2007.15967},
       adsurl = {https://ui.adsabs.harvard.edu/abs/2021JPlPh..87a2001P}
}

@ARTICLE{Tabone2022,
       author = {{Tabone}, Beno{\^\i}t and {Rosotti}, Giovanni P. and {Cridland}, Alexander J. and {Armitage}, Philip J. and {Lodato}, Giuseppe},
        title = "{Secular evolution of MHD wind-driven discs: analytical solutions in the expanded {\ensuremath{\alpha}}-framework}",
      journal = {\mnras},
         year = 2022, volume = {512}, pages = {2290-2309},
          doi = {10.1093/mnras/stab3442},
       eprint = {2111.10145},
       adsurl = {https://ui.adsabs.harvard.edu/abs/2022MNRAS.512.2290T}
}

@ARTICLE{WafflardFernandez2023,
       author = {{Wafflard-Fernandez}, Gaylor and {Lesur}, Geoffroy},
        title = "{Planet-disk-wind interaction: The magnetized fate of protoplanets}",
      journal = {\aap},
         year = 2023, volume = {677}, eid = {A70},
          doi = {10.1051/0004-6361/202245305},
       eprint = {2305.11784},
       adsurl = {https://ui.adsabs.harvard.edu/abs/2023A&A...677A..70W}
}

@ARTICLE{Hsu26,
       author = {{Hsu}, Chun-Yen and {Li}, Zhi-Yun and {Hu}, Xiao and {Tu}, Yisheng and {Lin}, Min-Kai},
        title = "{Dust and gas transport in substructured non-ideal MHD wind-launching discs with embedded planets}",
      journal = {\mnras},
         year = 2026,
        month = sep,
       volume = {551},
       number = {2},
          eid = {stag1479},
        pages = {stag1479},
          doi = {10.1093/mnras/stag1479},
archivePrefix = {arXiv},
       eprint = {2607.20158},
 primaryClass = {astro-ph.EP},
       adsurl = {https://ui.adsabs.harvard.edu/abs/2026MNRAS.551g1479H}
}

@ARTICLE{Canup2002,
       author = {{Canup}, Robin M. and {Ward}, William R.},
        title = "{Formation of the Galilean Satellites: Conditions of Accretion}",
      journal = {\aj},
         year = 2002,
        month = dec,
       volume = {124},
       number = {6},
        pages = {3404-3423},
          doi = {10.1086/344684},
       adsurl = {https://ui.adsabs.harvard.edu/abs/2002AJ....124.3404C}
}

%\begin{thebibliography}{99}
%\end{thebibliography}

%%%%%%%%%%%%%%%%%%%%%%%%%%%%%%%%%%%%%%%%%%%%%%%%%%

%%%%%%%%%%%%%%%%% APPENDICES %%%%%%%%%%%%%%%%%%%%%

%%%%%%%%%%%%%%%%%%%%%%%%%%%%%%%%%%%%%%%%%%%%%%%%%%

% Don't change these lines
\bsp	% typesetting comment
\label{lastpage}
\end{document}